\documentclass[10pt,twocolumn,twoside]{IEEEtran}
\usepackage{ifpdf}
\usepackage{url}
\ifpdf
\else
\fi
\usepackage{acro}
\usepackage{cite}
\usepackage{balance}
\usepackage{bm,comment,color}

\ifCLASSINFOpdf
  \usepackage[pdftex]{graphicx}
  \graphicspath{{../pdf/}{../jpeg/}}
  \DeclareGraphicsExtensions{.pdf,.jpeg,.png}
\else
  \usepackage[dvips]{graphicx}
  \graphicspath{{../eps/}}
  \DeclareGraphicsExtensions{.eps}
\fi
\usepackage{amsmath}
\usepackage{algpseudocode}
\algtext*{EndWhile}
\algtext*{EndIf}
\algtext*{EndFor}
\usepackage[ruled,vlined]{algorithm2e} 
\newcommand{\Input}[1]{\textbf{Input:} #1\\}
\newcommand{\Output}[1]{\textbf{Output:} #1\\}
\usepackage{array}
\usepackage{amsfonts} 
\usepackage{amssymb}
\usepackage{esint} 
\usepackage{units}
\usepackage{multirow}
\usepackage{comment}

\usepackage{bbm}

\usepackage{color}

\usepackage{standalone}

\usepackage{pgfplots}
\usepackage{tikz}
\usetikzlibrary{calc}
\makeatletter
\newcommand{\gettikzxy}[3]{%
  \tikz@scan@one@point\pgfutil@firstofone#1\relax
  \edef#2{\the\pgf@x}%
  \edef#3{\the\pgf@y}%
}
\usetikzlibrary{spy,backgrounds}
\pgfplotsset{compat=newest}
\usetikzlibrary{plotmarks}
\usetikzlibrary{arrows.meta}
\usepgfplotslibrary{patchplots}
\usepackage{grffile}
\newlength\fheight 
\newlength\fwidth 
\usepgfplotslibrary{fillbetween}

\usepackage{tabu,longtable}

\ifCLASSOPTIONcompsoc
 \usepackage[caption=false,font=normalsize,labelfont=sf,textfont=sf]{subfig}
\else
 \usepackage[caption=false,font=footnotesize]{subfig}
\fi
\usepackage{stfloats}
\usepackage[hidelinks]{hyperref}
\usepackage{xcolor}
\long\def\comment#1{}

\newfont{\bbb}{msbm10 scaled 700}

\newfont{\bb}{msbm10 scaled 1100}

\allowdisplaybreaks

\usepackage{graphicx}      
\renewcommand{\baselinestretch}{0.985} 
\usepackage{booktabs}

\usepackage{amsthm}
\newtheorem{theorem}{Theorem}
\usepackage{titlesec}
\titlespacing{\subsection}{0pt}{*0.5}{*0.3}
\titlespacing{\section}{0pt}{*0.5}{*0.3}

\renewcommand{\thesubsubsection}{\arabic{subsubsection})}
\makeatletter
\@addtoreset{subsubsection}{subsection}
\makeatother
\titleformat{\subsubsection}[runin]
  {\normalfont\normalsize\itshape}      %
  {\quad \thesubsubsection}          %
  {0.5em}                       %
  {}[:\hspace{0.5em}]                         %
\titlespacing{\subsubsection}{0pt}{*0.5}{1em} %

\usepackage{caption}
\begin{document} 
\bstctlcite{IEEEexample:BSTcontrol}

\title{Large Wireless Localization Model (LWLM): A Foundation Model for Positioning in 6G Networks}


\author{
Guangjin Pan, Kaixuan Huang, Hui Chen, \IEEEmembership{Member, IEEE}, Shunqing Zhang, \IEEEmembership{Senior Member, IEEE}, \\
Christian Häger, \IEEEmembership{Member, IEEE}, Henk Wymeersch, \IEEEmembership{Fellow, IEEE}
\thanks{G. Pan, H. Chen, C. Häger, and H. Wymeersch are with Department of Electrical Engineering, Chalmers University of Technology, 41296 Gothenburg, Sweden (e-mail: {guangjin.pan; hui.chen; christian.haeger; henkw}@chalmers.se)}

\thanks{K. Huang, S. Zhang, are  with the School of Communication and Information Engineering, Shanghai University, Shanghai
 200444, China.
Shanghai (e-mail: {xuan1999;  shunqing}@shu.edu.cn)}

\thanks{This work was supported in part by a grant from the Chalmers AI Research
 Center Consortium (CHAIR), by the National Academic Infrastructure for
 Supercomputing in Sweden (NAISS), by the SNS JU project 6G-DISAC under the EU's Horizon Europe research and innovation Program under Grant Agreement No. 101139130, the Swedish Foundation for Strategic Research (SSF) (grant FUS21-0004, SAICOM),  and Chalmers Areas of Advance in ICT and Transport. The work of C.~Häger was supported by the Swedish Research Council under grant No.~2020-04718.
}
}


\maketitle

\begin{abstract}
Accurate and robust localization is a critical enabler for emerging 5G and 6G applications, including autonomous driving, extended reality (XR), and smart manufacturing. While data-driven approaches have shown promise, most existing models require large amounts of labeled data and struggle to generalize across deployment scenarios and wireless configurations. To address these limitations, we propose a foundation-model-based solution tailored for wireless localization. We first analyze how different self-supervised learning (SSL) tasks acquire general-purpose and task-specific semantic features based on information bottleneck (IB) theory. Building on this foundation, we design a pretraining methodology for the proposed Large Wireless Localization Model (LWLM). Specifically, we propose an SSL framework that jointly optimizes three complementary objectives: (i) spatial-frequency masked channel modeling (SF-MCM), (ii) domain-transformation invariance (DTI), and (iii) position-invariant contrastive learning (PICL). These objectives jointly capture the underlying semantics of wireless channel from multiple perspectives. We further design lightweight decoders for key downstream tasks, including time-of-arrival (ToA) estimation, angle-of-arrival (AoA) estimation, single base station (BS) localization, and multiple BS localization. Comprehensive experimental results confirm that LWLM consistently surpasses both model-based and supervised learning baselines across all localization tasks. In particular, LWLM achieves 26.0\%--87.5\% improvement over transformer models without pretraining, and exhibits strong generalization under label-limited fine-tuning and unseen BS configurations, confirming its potential as a foundation model for wireless localization.
\end{abstract}


\IEEEpeerreviewmaketitle
\acresetall 

\section{Introduction}
As wireless systems evolve beyond basic communication functions, precise and reliable localization has become an essential capability, playing a crucial role in 5G and upcoming 6G networks \cite{Wymeersch20256G}. With location information, a wide array of applications can be supported, including autonomous vehicles, unmanned aerial vehicle (UAV) coordination, and extended reality (XR) \cite{Italiano2024Tutorial}. Accurate location information also enhances system-level functionalities such as mobility management \cite{di2014location}, resource scheduling \cite{kwon2023integrated}, and interference mitigation \cite{chen2024location}.
%
Wireless localization is usually addressed using model-based approaches, including time-of-arrival (ToA), time-difference-of-arrival (TDoA), angle-of-arrival (AoA), and received signal strength (RSS) measurements \cite{chen2022tutorial}. Classical algorithms such as multiple signal classification (MUSIC) and orthogonal matching pursuit (OMP) can estimate distance and angle between the user equipment (UE) and the base station (BS) for localization \cite{keskin2021mimo}. However, model-based methods degrade in complex environments such as multipath-rich and non-line-of-sight (NLoS) scenarios, especially under limited infrastructure. Therefore, achieving high-accuracy localization in dynamic, complex wireless environments is a key goal for future communication systems. To overcome these limitations, artificial intelligence (AI)-based localization \cite{pan2025ai} has emerged as a promising technique, allowing AI models to learn direct mappings from channel state information (CSI) to position without the need for explicit channel modeling.

\subsection{Related Works}

Data-driven supervised learning methods train models to map CSI to user locations using large-scale labeled datasets. To enhance the model’s fitting ability and accuracy, various neural network architectures have been studied, including multilayer perceptrons (MLPs) \cite{gao2023metaloc}, convolutional neural networks (CNNs) \cite{wu2021learning,pan2020High}, and long short-term memory (LSTM) networks \cite{chen2023deep, klus2024robust}. With rapid advances in AI techniques, transformers, which have revolutionized fields such as natural language processing (NLP) and computer vision (CV), are also showing their strong potential in localization tasks \cite{liu2022transformer, gong2023deep, li2023lot, xu2024swin}. Nevertheless, most of these supervised learning models are task-specific and require a large amount of labeled data for each new deployment scenario. Their heavy sample dependence leads to high deployment overhead and poor generalization across diverse environments and configurations.

To reduce the dependency on labeled data, semi-supervised and unsupervised learning approaches have been proposed. Some studies adopt autoencoders or generative adversarial networks (GANs) to reconstruct or simulate CSI without relying on ground-truth positions. For instance, iPos-5G~\cite{5G2023Ruan} uses an autoencoder to extract CSI features for indoor localization. In \cite{Chen2022Fidora}, the authors leverage variational autoencoder (VAE)-based data augmentation and joint classification-reconstruction networks for domain-adaptive localization. The authors in~\cite{Junoh2024Enhancing} further propose a semi-crowdsourced radio map construction framework using GAN-based augmentation to reduce annotation effort. Other works apply transfer learning and domain adaptation to reduce distribution shifts between source and target domains \cite{li2021transloc}. While these methods alleviate labeling costs to some extent, they typically do not learn robust, location-dependent, and transferable representations. In particular, they fail to capture high-level semantic features, like spatial relationships or environment-dependent channel patterns, which can be exploited to support a wide range of channel-related tasks and improve their performance. As a result, their performance often deteriorates in heterogeneous scenarios, such as varying bandwidths and pilot configurations.

In recent years, the foundation model paradigm~\cite{bommasani2021opportunities} using Large AI models (LAMs) has demonstrated transformative success in domains such as NLP, CV, and audio signal processing. These models are pretrained on large-scale, unlabeled datasets using self-supervised learning (SSL) \cite{gui2024survey} to acquire general-purpose, transferable representations. Examples include BERT~\cite{devlin2019bert} and GPT~\cite{brown2020language} in NLP, as well as SpectralGPT~\cite{Hong2024SpectralGPT} for audio-based tasks. These models serve as \textit{universal encoders for diverse downstream applications} and have shown significant improvements in semantic understanding, domain transferability, and training efficiency for downstream fine-tuning~\cite{fei2022towards, He2024DownStream}.
Based on this paradigm, wireless communications can also benefit from the foundation model paradigm to overcome the problems of scarce labeled data and insufficient generalization capabilities for scenarios, environments, and configurations \cite{alikhani2024large, yang2025wirelessgpt,Liu2024WiFo,liu2024llm4cp,liu2025llm4wm}. A well-designed wireless foundation model can serve as a universal semantic encoder, capturing physical propagation characteristics that generalize across deployments, hardware, and tasks. By formulating appropriate SSL tasks over large-scale unlabeled CSI datasets, SSL enables neural networks to discover the latent semantics of wireless channels. Combined with transformer and well-designed SSL tasks~\cite{gui2024survey}, such models have the potential to serve as wireless foundation models, and can be fine-tuned for a wide range of downstream tasks.

SSL algorithms mainly include generative-based and contrastive-based approaches~\cite{gui2024survey}. Building on these techniques, several recent works have explored wireless foundation models for general channel-related tasks. Based on a generative-based approach, LWM~\cite{alikhani2024large} applies masked reconstruction in the spatial-frequency domain for downstream tasks such as channel estimation and beamforming. WirelessGPT~\cite{yang2025wirelessgpt} extends masked modeling to three-dimensional (3D) channel representations across spatial, time, and frequency, achieving superior performance in channel estimation, prediction, human activity recognition, and environment reconstruction. WiFo~\cite{Liu2024WiFo} unifies frequency- and time-domain channel prediction under a masked SSL formulation and exhibits strong zero-shot generalization. Based on a contrastive-based approach, the authors in~\cite{Salihu2024Self-Supervised} employ data augmentation techniques (e.g., random subcarrier selection and random subcarrier flipping) to construct positive samples, enabling the learning of invariant representations of wireless channels. In addition, LLM4CP~\cite{liu2024llm4cp} and LLM4WM~\cite{liu2025llm4wm} directly fine-tune large language models (LLMs) for wireless channel tasks, transferring language representations to channel representations. Although these works demonstrate the feasibility and effectiveness of foundation models in wireless tasks and may perform well in localization scenarios, none of them is specifically designed for wireless localization. Moreover, contrastive learning has also been employed to capture the correlations between channel samples for low-dimensional channel charting \cite{stephan2024angle}. While such charting effectively visualizes the similarities among channel samples, it lacks the generalizable representations required for diverse downstream tasks.

In addition, some prior works have proposed SSL-based pretraining frameworks for localization. For example, CrowdBERT~\cite{Han2024CrowdBERT} applies a masking strategy over RSS fingerprints from multiple access points and fine-tunes the model with limited labeled data. In~\cite{Wang2025Signal}, a signal-guided masked autoencoder uses antenna-domain masking and channel attention to reconstruct channel impulse responses (CIRs) for multi-BS localization. In~\cite{ott2024radio}, the paper reconstructs masked CIRs to extract environment-specific semantics, enabling accurate single-BS localization. However, these methods \cite{Han2024CrowdBERT, Wang2025Signal, ott2024radio} target specific localization tasks, lacking generalization across tasks (e.g., AoA/ToA estimation, single/multi-BS localization) and different BS configurations. Moreover, these methods typically adopt a single SSL pretraining objective, limiting the diversity of the learned channel semantics.

\subsection{Contributions}

In this paper, we propose a \textit{Large Wireless Localization Model (LWLM)} \footnote{The code is available at: https://github.com/guangjinpan/LWLM.}, a self-supervised foundation model specifically tailored for wireless localization. LWLM is designed to address the key challenges in generalization, adaptability, and data efficiency across diverse localization tasks and deployment configurations. The main contributions are as follows:
\begin{itemize}
    \item \textbf{Information bottleneck analysis for SSL in foundation model:} We establish an analysis framework for SSL by adopting the information bottleneck (IB) theory. Through this framework, we analyze the underlying principles of generative-based SSL in suppressing noise and redundancy to obtain general-purpose semantic representations. We also discuss how contrastive-based SSL indirectly optimizes the IB objective by sampling different samples with the same label, thereby capturing localization-relevant semantic features. These insights can inform the design of effective pretraining tasks for wireless foundation models.

     \item \textbf{Hybrid self-supervised learning framework for LWLM:} Based on the conclusions from the IB framework, we propose a novel hybrid SSL pretraining framework for wireless localization tasks that jointly optimizes three objectives: 
    (i) spatial-frequency masked channel modeling (SF-MCM) to capture local and global correlations across antennas and subcarriers 
    (ii) domain-transformation invariance (DTI) for enforcing consistency across spatial-frequency and angle-delay domains, and 
    (iii) position-invariant contrastive learning (PICL) for capturing robust, location-dependent semantics invariant to BS configurations. While SF-MCM builds upon existing masked modeling techniques, DTI and PICL are newly introduced.
    This joint learning framework enables the model to acquire diverse, configuration-invariant, and location-aware representations from unlabeled CSI.

    \item \textbf{Task-adaptive decoders for downstream localization tasks:} We develop lightweight task-specific decoders for key downstream localization tasks, including ToA estimation, AoA estimation, single-BS localization, and multi-BS fusion. To support an arbitrary number of BSs, the multi-BS decoder extends single-BS models with an attention-based fusion module that adaptively aggregates the location estimates of each BS.

\end{itemize}
Experiments demonstrate that LWLM consistently achieves state-of-the-art performance across all tasks compared to model-based and supervised-learning-based baselines. Specifically, compared with the transformer without pretraining, LWLM achieves improvements of 26.0\%, 36.7\%, 87.5\%, and 32.0\% in ToA estimation, AoA estimation, single-BS localization, and multi-BS localization tasks, respectively. Moreover, LWLM demonstrates strong performance in label-limited scenarios, generalizes well to unseen deployment configurations, and remains compatible with pilot-based CSI measurements for practical communication systems.


\section{System Model}
\label{sec:SystemModel}

Considering an uplink multiple-input multiple-output orthogonal frequency-division multiplexing (MIMO-OFDM) system, a generic UE is equipped with a single omnidirectional antenna, while a generic BS is equipped with a uniform linear array (ULA) consisting of $N_{\text{ant}}$ antennas. The total system bandwidth, denoted by $B_{\text{bw}}$, is uniformly divided into $N_{\text{subc}}$ orthogonal subcarriers, each with bandwidth $\Delta_f={B_{\text{bw}}}/{N_{\text{subc}}}$.

\subsection{Channel Model}

Let $\bm{p}^{\text{bs}} \in \mathbb{R}^{d_p}$ and $\bm{p}^{\text{ue}} \in \mathbb{R}^{d_p}$ denote the spatial coordinates of the BS and UE, respectively, where $d_p = 2$ indicates a two-dimensional localization system. The CSI vector at the $k$-th subcarrier, denoted by $\bm{h}_{k} \in \mathbb{C}^{N_{\text{ant}}}$, characterizes the uplink channel from the UE to the BS and is modeled via the channel frequency response (CFR) as 
    $\bm{h}_{k} = \sum_{l=1}^{L} \alpha_l \bm{a}^{\text{bs}}(\theta_l) e^{-j 2\pi k\Delta_f \tau_l} + \bm{n}_{k}$, 
where $L$ is the number of multipath components (MPCs), $\alpha_l \in \mathbb{C}$ is the complex gain of the $l$-th MPC, $\tau_l$ denotes its propagation delay, and $\theta_l$ is the angle of arrival (AoA) at the BS. $\bm{n}_{k} \in \mathbb{C}^{N_{\text{ant}}}$ represents the additive white Gaussian noise (AWGN) vector. The ULA array response vector $\bm{a}^{\text{bs}}(\cdot) \in \mathbb{C}^{N_{\text{ant}}}$ for a given AoA $\theta$ is formulated as
$\bm{a}^{\text{bs}}(\theta) = [1, e^{-j \frac{2\pi d}{\lambda}\sin(\theta)}, \dots, e^{-j \frac{2\pi d}{\lambda}(N_{\text{ant}} - 1)\sin(\theta)}]^\top$, 
where $\lambda$ and $d$ denote the carrier wavelength and the antenna spacing, respectively.
Aggregating CFRs from all subcarriers, the CFR matrix 
is 
$\bm{H} = [\bm{h}_{1}, \dots, \bm{h}_{N_{\text{subc}}}] \in \mathbb{C}^{N_{\text{ant}} \times N_{\text{subc}}}$.

\begin{figure}
\centering
\includegraphics[scale =0.40]{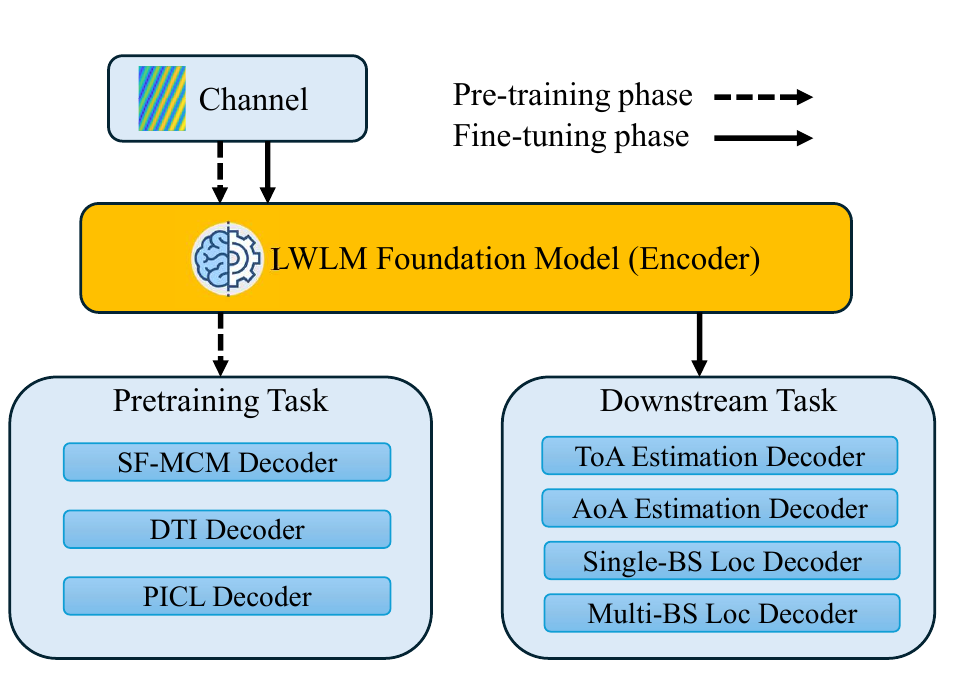}
\caption{Schematic diagram of the LWLM algorithm framework. First, the LWLM foundation model is pretrained using the pretraining task, and then fine-tuned on specific localization tasks.}
\label{fig:LWLMFramework} 
\end{figure}

\subsection{Problem Formulation}

Learning robust, generalizable representations from channel measurements remains a key challenge in AI-driven wireless localization~\cite{pan2025ai}. To tackle this, we propose a general-purpose channel representation framework based on SSL, named LWLM (see Fig.~\ref{fig:LWLMFramework}).
Specifically, LWLM is a transformer-based encoder designed to learn universal and transferable channel representations from a large-scale unlabeled channel dataset, denoted as $\mathcal{D}^{\text{pre}} = \{\bm{H}_s | s = 1, \dots, N^{\text{pre}}\}$, where $N^{\text{pre}}$ is the total number of pretraining samples. The LWLM encoder $\mathcal{F}_{\bm{\Phi}}^{\text{LWLM}}(\cdot)$, parameterized by $\bm{\Phi}$, transforms input CFR sample $s$ into a latent representation:
$\bm{o}_s = \mathcal{F}_{\bm{\Phi}}^{\text{LWLM}}(\bm{H}_s)$, 
where $\bm{o}_s$ is the channel semantics, which includes the essential propagation properties of the wireless channel in a task-independent manner. To train the LWLM encoder $\mathcal{F}_{\bm{\Phi}}^{\text{LWLM}}(\cdot)$ and obtain general-purpose representations $\bm{o}_s$ beneficial for localization-related tasks, we design hybrid SSL tasks for the pretraining stage. Let $\mathcal{K}_{\text{s-task}} = {1, 2, \dots, K_{\text{s-task}}}$ denote the set of $K_{\text{s-task}}$ SSL pretraining tasks. Each pretraining task $k_{\text{s-task}}$ has an associated decoder $\mathcal{F}_{\bm{\Theta}_{k_{\text{s-task}}}}^{\text{s-task}}$, and the overall optimization objective for the pretraining is formulated as
\begin{align}
\min_{\bm{\Phi}, \{\bm{\Theta}_{k_{\text{s-task}}}\}} \quad & \! \! \! \! \! \! \sum_{k_{\text{s-task}} \in \mathcal{K}_{\text{s-task}}}  \! \! \! \! \! \! \alpha_{k_{\text{s-task}}} \mathbb{E}_{\bm{H}_s \sim \mathcal{D}^{\text{pre}}} \! \! \left[\mathcal{L}_{k_{\text{s-task}}}\left(\mathcal{F}_{\bm{\Theta}_{k_{\text{s-task}}}}^{\text{s-task}}(\bm{o}_s), \bm{H}_s \right)\right], \nonumber \\
\text{s.t.} \quad & \bm{o}_s = \mathcal{F}_{\bm{\Phi}}^{\text{LWLM}}(\bm{H}_s),
\end{align}
where $\mathcal{L}_{k_{\text{s-task}}}(\cdot,\cdot)$ is the loss function associated with SSL task $k_{\text{s-task}}$, and $\alpha_{k_{\text{s-task}}} \ge 0$ is its corresponding weight.

After the pretraining, the pretrained LWLM encoder is adapted to multiple downstream localization tasks through lightweight, task-specific decoders. Let $\mathcal{K}_{\text{d-task}} = \{1, 2, \dots, K_{\text{d-task}}\}$ denote the set of downstream tasks. For each localization-related task ${k_{\text{d-task}}} \in \mathcal{K}_{\text{d-task}}$, we define a decoder $\mathcal{F}_{\bm{\Theta}_{k_{\text{d-task}}}}^{\text{d-task}}$ with parameters $\bm{\Theta}_{k_{\text{d-task}}}$. The model is fine-tuned using a small labeled dataset $\mathcal{D}^{\text{d-task}}_{k_{\text{d-task}}} = \{(\bm{H}_s, \bm{y}_s^{k_{\text{d-task}}}) | s = 1, 2, \dots, N^{\text{d-task}}_{k_{\text{d-task}}}\}$, where $\bm{y}_s^{k_{\text{d-task}}}$ denotes the task-specific label (e.g., ToA, AoA, 2D coordinates with $d_p = 2$), and $N^{\text{d-task}}_{k_{\text{d-task}}}$ is the number of labeled samples for task $k_{\text{d-task}}$. The fine-tuning objective for each downstream task $k_{\text{d-task}}$ can be expressed as
\begin{align}
\min_{\bm{\Phi}, \bm{\Theta}_{k_{\text{d-task}}}} \quad & \mathbb{E}_{(\bm{H}_s, \bm{y}^{k_{\text{d-task}}}_s) \sim \mathcal{D}^{\text{d-task}}_{k_{\text{d-task}}}} \! \! \left[\mathcal{L}_{k_{\text{d-task}}}\left(\mathcal{F}_{\bm{\Theta}_{k_{\text{d-task}}}}^{\text{d-task}}(\bm{o}_s), \bm{y}^{k_{\text{d-task}}}_s \right)\right], \nonumber\\
\text{s.t.} \quad & \bm{o}_s = \mathcal{F}_{\bm{\Phi}}^{\text{LWLM}}(\bm{H}_s),
\end{align}
where $\mathcal{L}_{k_{\text{down}}}(\cdot, \cdot)$ is the task-specific loss function for task $k_{\text{down}}$. Notably, during the fine-tuning phase, the parameters $\bm{\Phi}$ of the LWLM encoder can be either frozen or updated, depending on factors such as the similarity between the downstream and pretraining tasks, the amount of labeled data, and the complexity of the downstream decoder. In this work, since the downstream task differs significantly from the pretraining objective and we adopt a lightweight decoder, we fine-tune the encoder during downstream training.

\color{black}

\section{Information Bottleneck Analysis for SSL}
\label{sec:IB}

To gain theoretical insights into the design of SSL pretraining objectives, we adopt the IB framework~\cite{hu2024survey} as an analytical tool to characterize the trade-off between generalization and task relevance in representation learning. The IB objective aims to optimize the encoder and decoder by minimizing
\begin{align}
\mathcal{L}_{\text{IB}}(\bm{\Phi},\bm{\Theta}) = I(O;H) - \beta I(O;Y),
\label{eq_loss_IB}
\end{align}
where $H$, $O$, and $Y$ denote the random variables corresponding to the input channel matrix, the learned intermediate representation, and the downstream task labels, respectively. Their realizations are denoted by $\bm{H}$, $\bm{o}$, and $\bm{y}$. The weighting factor $\beta > 0$ controls the trade-off between compression and prediction, and $\bm{\Phi}$ and $\bm{\Theta}$ are the trainable parameters of the encoder and decoder, respectively.
To minimize the IB objective, the first term encourages the encoder to discard task-irrelevant noise and redundant information from the input, while the second term encourages the decoder to retain task-relevant features from the learned representation $O$. The trade-off between these two objectives governs the balance between generality and specificity in representation learning. Although this supervised formulation provides an ideal framework for learning meaningful semantic representations, it is often impractical to directly optimize in real-world settings due to factors such as limited labeled data, model capability constraints, and generalization issues.

In the context of foundation models, we typically expect the learned representation $O$ to encode semantics that are transferable to multiple downstream tasks. This requires SSL objectives that remove irrelevant channel information (such as noise) while preserving as much general-purpose semantic content as possible. Based on this principle, we next analyze how different types of SSL objectives can support this goal within the IB framework. In particular, we consider two types of SSL tasks: \textit{generative-based SSL} and \textit{contrastive-based SSL} tasks. First, we analyze how generative-based SSL methods can promote generalization by encouraging the model to reconstruct underlying channel structures. At the same time, since our primary goal is to enable effective wireless localization, we further analyze how contrastive-based SSL can help tailor the representation $O$ to be more discriminative and relevant to localization tasks.

\subsection{Generative-based SSL within IB Framework}
As mentioned above, to enable the learned representation $O$ to discard semantically irrelevant information (e.g., channel estimation noise) and prevent overfitting while ensuring generalization across diverse downstream tasks, our goal is to minimize the mutual information $I(O;H)$. However, directly minimizing $I(O;H)$ may lead to results where $O$ and $H$ become completely independent, thus losing useful semantic information. To address this issue, we introduce generative-based SSL tasks to preserve the channel semantics. Specifically, we consider a set of generative-based SSL tasks involving channel transformations $\{T_{k_g}(H)\}_{k_g=1}^{K_g}$ (e.g., masked modeling, domain transformation), where $K_g$ denotes the number of generative-based SSL tasks used during the pretraining stage. For each SSL task $k_g$, the corresponding IB objective is formulated as
\begin{align}
\mathcal{L}_{\text{IB-G}}(\bm{\Phi},\bm{\Theta}) = I(O;H) - \beta_{k_g} I(O;T_{k_g}(H)),
\end{align}
where $\beta_{k_g}$ is the weighting factor, and is generally greater than 1, as indicated by later derivations. Based on this, we present the following theorem (proof in Appendix \ref{proof:theorem1}):

\begin{theorem}
\label{theorem:1}
\vspace{-2mm}
For the transformation $T_{k_g}(H)$, let $\overline{T}_{k_g}(H)$ denote the residual information in $H$ after extracting $T_{k_g}(H)$, such that $(T_{k_g}(H),\,\overline{T}_{k_g}(H))$ is bijective with $H$.
Define the reconstruction loss for $T_{k_g}(H)$ and residual reconstruction loss for $\overline{T}_{k_g}(H)$ for the generative-based SSL task $k_g$ as
\begin{align}
     \mathcal{L}_{k_g}\! (\bm{\Phi},\bm{\Theta}_{k_g}) \! = \! & -\! \mathbb{E}_{p(H)q_{\bm{\Phi}}(O|H)}\! \! \left[\ln p_{\bm{\Theta}_{k_g}}\! \left(T_{k_g}(H)\,|\,O\right)\! \right] \nonumber \\
     \overline{\mathcal{L}}_{k_g}\! (\bm{\Phi},\overline{\bm{\Theta}}_{k_g}) \! = \! &  - \! \mathbb{E}_{p(H)q_{\bm{\Phi}}(O|H)}\! \left[\ln p_{\overline{\bm{\Theta}}_{k_g}}\! \! \left(\overline{T}_{k_g}(H)|T_{k_g}(H),O\right)\!  \right] \nonumber
\end{align}
where $\bm{\Phi}$ are the parameters of the encoder, ${\bm{\Theta}}_{k_g}$ are the parameters of the decoder for task-relevant output $T_{k_g}$, and $\overline{\bm{\Theta}}_{k_g}$ are the parameters of the decoder for the residual component $\overline{T}_{k_g}$.
Then, the IB objective $\mathcal{L}_{\text{IB-G}}(\bm{\Phi}, \bm{\Theta})$ can be approximated as
\begin{align}
\mathcal{L}_{\text{IB-G}}(\bm{\Phi},\bm{\Theta}) \approx (\beta_{k_g}-1) \mathcal{L}_{k_g}(\bm{\Phi},\bm{\Theta}_{k_g}) -  \overline{\mathcal{L}}_{k_g}(\bm{\Phi},\overline{\bm{\Theta}}_{k_g}) + C. \nonumber
\end{align}
$(\beta_{k_g}-1)>0$ denotes the trade-off between preserving task-relevant information and discarding residual information, and $C$ is a constant independent of parameters $\bm{\Phi}$, $\bm{\Theta}_{k_g}$ and $\overline{\bm{\Theta}}_{k_g}$.
\end{theorem}

Based on Theorem \ref{theorem:1}, the first term $\mathcal{L}_{k_g}(\bm{\Phi},\bm{\Theta}_{k_g})$ encourages to learn a better representation $O$ to accurately reconstruct $T_{k_g}(H)$, while the negative second term $\overline{\mathcal{L}}_{k_g}(\bm{\Phi},\overline{\bm{\Theta}}_{k_g})$ encourages $O$ to suppress the reconstruction of $\overline{T}_{k_g}(H)$, hence discarding residual information. This highlights a trade-off in the representation $O$ to preserve important semantics from ${T}_{k_g}(H)$ and discard irrelevant or noisy components. It is worth noting that, in the actual training process, $\overline{T}_{k_g}(H)$ is typically not directly accessible. As a result, low-dimensional representations of $O$, along with dropout and normalization layers, are often employed to help the neural network implicitly discard residual information. Furthermore, we generally set $(\beta_{k_g}-1)>0$ to encourage the AI model to learn the important channel semantics while reconstructing ${T}_{k_g}(H)$. Based on Theorem \ref{theorem:1}, we can infer that incorporating a richer set of generative-based SSL tasks $\{T_{k_g}(H)\}$ can further guide the representation $O$ to minimize shared redundant information across these tasks (such as noise). This, in turn, will result in $O$ containing richer and more meaningful semantics of the channel.
However, because these tasks do not incorporate downstream-specific signals, the resulting representations may lack optimal task specificity. In the next section, we explain how contrastive-based SSL can introduce targeted downstream semantics to further enhance task-specific performance.

\subsection{Contrastive-based SSL within IB Framework}

To extract task-dependent semantics tailored for downstream localization tasks, our goal is to let channel representation $O$ include more location-related semantics during pretraining. Since ground-truth labels $\bm{y}$ are typically unavailable during pretraining, we employ a contrastive learning method to indirectly obtain localization-related semantics. For localization, measurements from the same location under varying wireless conditions form positive pairs, while those from different locations serve as negative pairs. To illustrate that contrastive-based SSL can extract task-related semantics, we first give the following theorem (proof in Appendix \ref{proof:theorem2}):
\begin{theorem} 
\vspace{-2mm}
\label{theorem:2}
Let $\mathcal{N}_{\text{bat}} = \{(\bm{H}_s, \bm{y}_s)\}_{s=1}^{N_{\text{bat}}} \subset \mathcal{D}^{\text{pre}}$ be a batch of samples drawn independently from the joint distribution $p(H,Y)$, where $N_{\text{bat}}$ denotes the batch size. For each sample $s$ with target label $\bm{y}_s$, we additionally sample another channel realization $\bm{H}_{s^+} \sim p(H \mid Y = \bm{y}_s)$, which serves as a positive sample. The corresponding representations are obtained by $\bm{o}_s \sim q_{\bm{\Phi}}(\bm{o}|\bm{H}_s)$, and $\bm{o}_{s^+} \sim q_{\bm{\Phi}}(\bm{o}|\bm{H}_{s^+})$. 
Constructing the InfoNCE contrastive loss: 
\begin{align} 
\! \! \! L_{\text{InfoNCE}} = -\frac{1}{N_{\text{bat}}} \sum_{s=1}^{N_{\text{bat}}} \left[ \ln\frac{\exp\bigl(\mathrm{sim}(\bm{o}_s,\bm{o}_{s^+})/\tau\bigr)} {\sum \exp\bigl(\mathrm{sim}(\bm{o}_s,\bm{o}_j)/\tau\bigr)} \right], 
\end{align} where $\mathrm{sim}(\cdot,\cdot)$ denotes cosine similarity, and $\tau>0$ is a temperature parameter, the contrastive objective provides the following lower bound for $I(O;Y)$: 
\begin{align}  
I(O;Y) \ge I(O;O^+)  \ge \log(N_{\text{bat}})-L_{\text{InfoNCE}}, 
\end{align}
where the constant term is independent of $\bm{\Phi}$.
\end{theorem}

According to Theorem~\ref{theorem:2}, minimizing the InfoNCE loss in contrastive-based SSL is equivalent to maximizing a lower bound of $I(O;Y)$, thereby indirectly minimizing the original IB cost $I(H;O) - \beta I(O;Y)$. Therefore, contrastive-based SSL can capture the semantic representation of task dependencies. While this paper is primarily discussed in the context of localization, the underlying principle can be extended to other channel-related tasks by identifying meaningful invariances. Consequently, the proposed framework can be effectively adapted to a broad range of wireless applications, such as sensing and channel extrapolation.

In conclusion, generative-based SSL objectives help the model learn general-purpose semantic features beneficial to a variety of channel-related downstream tasks, as shown in prior works like \cite{alikhani2024large,yang2025wirelessgpt,Liu2024WiFo}. In contrast, contrastive-based SSL incorporates partial priors of target tasks during pretraining, producing more representations that perform better on specific tasks such as localization. This suggests that combining multiple SSL strategies enables the model to capture richer and more diverse channel semantics, leading to improved localization performance. Importantly, this insight also generalizes to other wireless applications, including sensing and communication.
\color{black}

\section{Self-Supervised Pretraining Framework}
\label{sec:SSL}

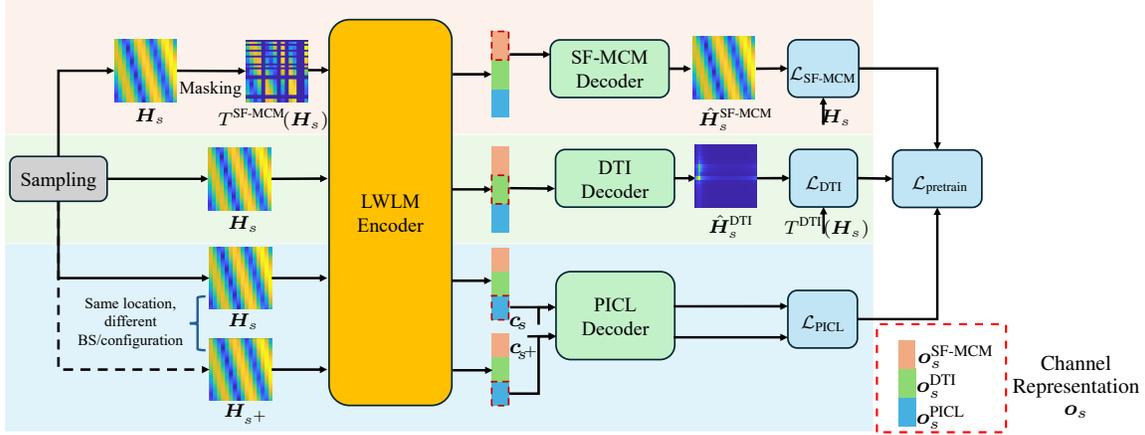
\begin{figure*}[t]
    \centering
    \begin{tikzpicture}
    \node (image) [anchor=south west]{\includegraphics[width=0.72\linewidth]{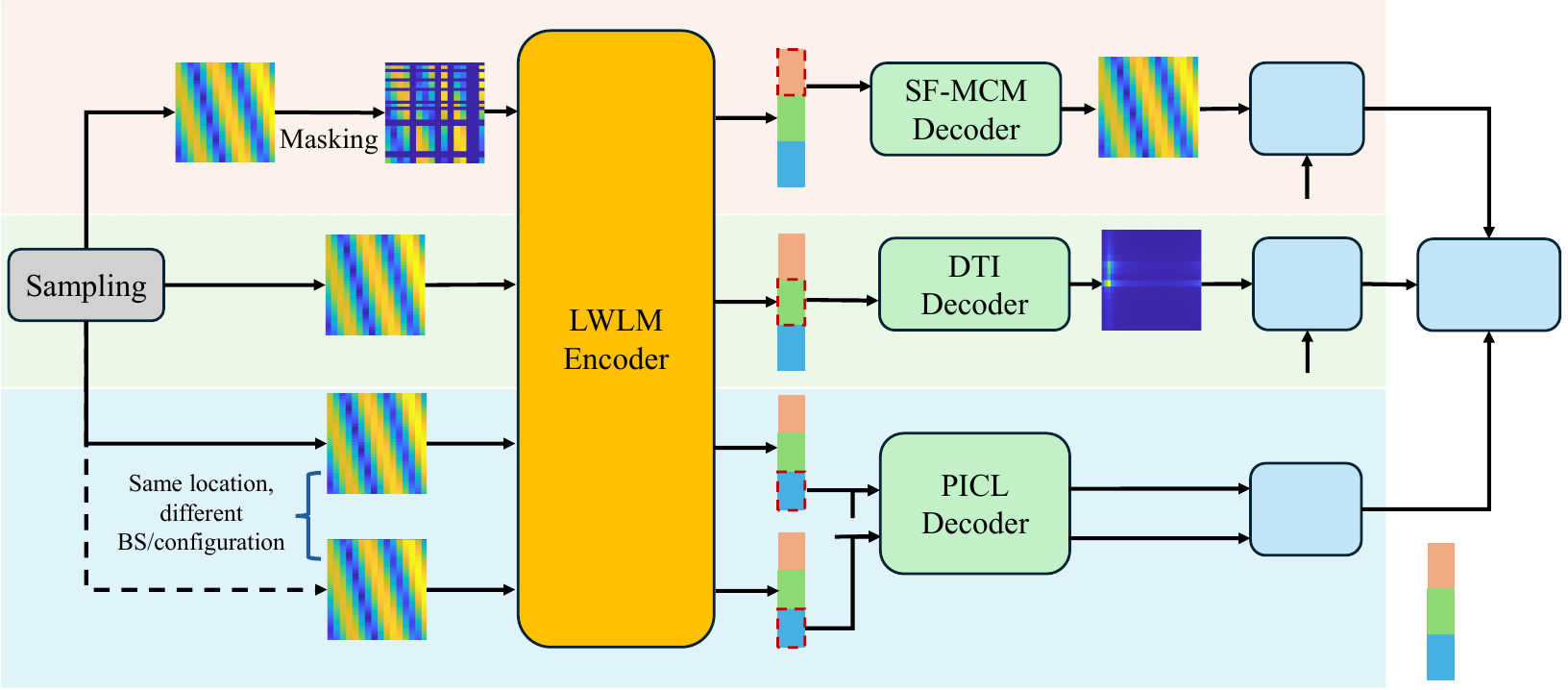}};
    \gettikzxy{(image.north east)}{\ix}{\iy};
    
    \node at (0.155*\ix,0.76*\iy)[rotate=0,anchor=north]{{\scriptsize $\bm{H}_s$}};
    \node at (0.278*\ix,0.76*\iy)[rotate=0,anchor=north]{{\scriptsize $T^{\text{SF-MCM}}\! (\bm{H}_s)$}};
    \node at (0.74*\ix,0.762*\iy)[rotate=0,anchor=north]{{\scriptsize $\hat{\bm{H}}_s^{\text{SF-MCM}}$}};
    \node at (0.84*\ix,0.757*\iy)[rotate=0,anchor=north]{{\scriptsize $\bm{H}_s$}};
    \node at (0.827*\ix,0.852*\iy)[rotate=0,anchor=north]{\scriptsize $\mathcal{L}_{\text{SF-MCM}}$};
    
    \node at (0.248*\ix,0.52*\iy)[rotate=0,anchor=north]{{\scriptsize $\bm{H}_s$}};
    \node at (0.735*\ix,0.523*\iy)[rotate=0,anchor=north]{{\scriptsize $\hat{\bm{H}}^{\text{DTI}}_s$}};
    \node at (0.831*\ix,0.518*\iy)[rotate=0,anchor=north]{{\scriptsize ${T}^{\text{DTI}} (\bm{H}_s)$}};
    \node at (0.828*\ix,0.61*\iy)[rotate=0,anchor=north]{\scriptsize$ \mathcal{L}_{\text{DTI}}$};
    
    \node at (0.248*\ix,0.304*\iy)[rotate=0,anchor=north]{{\scriptsize $\bm{H}_{s}$}};
    \node at (0.251*\ix,0.103*\iy)[rotate=0,anchor=north]{{\scriptsize $\bm{H}_{s^+}$}};
    \node at (0.523*\ix,0.297*\iy)[rotate=0,anchor=north]{{\footnotesize $\bm{c}_{\!s}$}};
    \node at (0.529*\ix,0.237*\iy)[rotate=0,anchor=north]{{\footnotesize $\bm{c}_{\!s^{\!+}}$}};
    \node at (0.827*\ix,0.30*\iy)[rotate=0,anchor=north]{\scriptsize $\mathcal{L}_{\text{PICL}}$};

    \node at (0.944*\ix,0.61*\iy)[rotate=0,anchor=north]{\scriptsize $\mathcal{L}_{\text{pretrain}}$};

    \node at (1.08*\ix,0.21*\iy)[rotate=0,anchor=north]{\footnotesize \shortstack {Channel \\ Representation \\ $\bm{o}_s$}};
    \node at (0.96*\ix,0.235*\iy)[rotate=0,anchor=north]{{\footnotesize $\bm{o}^{\text{SF-MCM}}_s$}};
    \node at (0.941*\ix,0.17*\iy)[rotate=0,anchor=north]{{\footnotesize $\bm{o}^{\text{DTI}}_s$}};
    \node at (0.945*\ix,0.1*\iy)[rotate=0,anchor=north]{{\footnotesize $\bm{o}^{\text{PICL}}_s$}};

    \draw[red, dashed, thick] (0.881*\ix, 0.02*\iy) rectangle (1.01*\ix, 0.260*\iy);
    
    \end{tikzpicture}
    \caption{Schematic diagram of the pretraining algorithm. The channel semantic representation $\bm{o}_s$ based on the hybrid SSL method is the concatenation of the three representations $\bm{o}^{\text{SF-MCM}_s}$, $\bm{o}^{\text{DTI}}_s$, $\bm{o}^{\text{PICL}}_s$. At each training step, we compute a weighted sum of three separate losses as the final pretraining loss.}
    
    \label{fig:system_setting}
    \vspace{-5mm}
\end{figure*}

Based on the above analysis, we develop a hybrid SSL framework tailored for wireless localization tasks. In particular, we identify and exploit three core properties of wireless channels and localization systems: (i) the spatial-frequency correlation across antennas and subcarriers, (ii) the invariance of channel information under transformations between different physical domains (e.g., spatial-frequency and angle-delay domains), and (iii) the position-level semantic invariance across varying network configurations. Based on these principles, we introduce three SSL strategies, i.e., SF-MCM, DTI, and PICL. The schematic diagram of the pretraining algorithm is shown in Fig. \ref{fig:system_setting}. Specifically, SF-MCM and DTI are generative-based SSL to obtain more general-purpose features, while PICL is contrastive-based SSL, enabling the model to learn task-relevant features even without direct access to location labels during pretraining. Additionally, SF-MCM is adapted from existing masked modeling works \cite{ alikhani2024large, yang2025wirelessgpt}, while DTI and PICL are newly introduced.

\subsection{SF-MCM-based SSL}
\label{SubSec:SFMCM}

SF-MCM is a generative SSL method that captures channel semantics by randomly masking spatial-frequency channel data and reconstructing it. This process encourages the encoder to learn both local and global contextual dependencies across antennas and subcarriers, while reducing redundancy.

Given a CFR sample $\bm{H}_s$, we randomly select subsets of antennas and subcarriers to be masked, denoted as $\mathcal{N}_{\text{ant}}^{M} \subseteq \{1,\dots, N_{\text{ant}}\}$ and $\mathcal{N}_{\text{subc}}^{M} \subseteq \{1,\dots, N_{\text{subc}}\}$, with $|\mathcal{N}_{\text{ant}}^{M}|=\hat{N}_{\text{ant}}^{M}$ and $|\mathcal{N}_{\text{subc}}^{M}|=\hat{N}_{\text{subc}}^{M}$, respectively. We define a binary masking matrix $\bm{M} \in \{0,1\}^{N_{\text{ant}} \times N_{\text{subc}}}$ as
\begin{align}
[M]_{i,j} = \begin{cases}
0, & i \in \mathcal{N}_{\text{ant}}^{M} \ or \ j \in \mathcal{N}_{\text{subc}}^{M} \\
1, & \text{otherwise}
\end{cases},
\end{align}
where $[M]_{i,j} = 0$ indicates that the corresponding spatial-frequency entry $(i,j)$ is masked (i.e., removed and requires reconstruction), while $[M]_{i,j} = 1$ denotes an unmasked entry.
The masked channel, denoted as $T^{\text{SF-MCM}}(\bm{H}_s)$, is given by
\begin{align}
    T^{\text{SF-MCM}}(\bm{H}_s) = \bm{H}_s \odot \bm{M},
    \label{equ:mask}
\end{align}
where $\odot$ denotes element-wise multiplication. The masked channel $T^{\text{SF-MCM}}(\bm{H}_s)$ serves as the input to the LWLM encoder to obtain a latent representation, i.e.,
$
    \bm{o}^{\text{SF-MCM}}_s = \mathcal{F}_{\bm{\Phi}}^{\text{LWLM}}(T^{\text{SF-MCM}}(\bm{H}_s))
$,
where $\bm{o}^{\text{SF-MCM}}_s$ represents the latent embedding learned from the SF-MCM pretraining objective.

Subsequently, the latent representation is decoded via a decoder $\mathcal{F}^{\text{SF-MCM}}_{\bm{\Theta}\text{SF-MCM}}(\cdot)$ to reconstruct the original CFR, i.e., $\hat{\bm{H}}_s^{\text{SF-MCM}} = \mathcal{F}^{\text{SF-MCM}}_{\bm{\Theta}_\text{SF-MCM}}(\bm{o}^{\text{SF-MCM}}_s)$, where $\hat{\bm{H}}_s^{\text{SF-MCM}}$ is the reconstructed channel by SF-MCM decoder.
As stated in Theorem~\ref{theorem:1}, minimizing the reconstruction error helps capture richer channel semantics. Therefore, we adopt the mean squared error (MSE) loss as the reconstruction objective:
\begin{align}
  \mathcal{L}_{\text{SF-MCM}} = \frac{1}{N_{\text{bat}}} \sum_{\bm{H}_s \in \mathcal{N}_{\text{bat}}} \left\| (\hat{\bm{H}}_s^{\text{SF-MCM}} - \bm{H}_s) \odot \bm{M} \right\|_F^2,
  \label{equ:LossSFMCM}
\end{align}
where $\|\cdot\|_F$ denotes the Frobenius norm.

\subsection{DTI-based SSL}
\label{SubSec:DTI}

DTI learns general-purpose features across different physically meaningful channel domains by learning their transformations. In this work, we consider the spatial-frequency and angle-delay domains, though the approach is extendable to others. By learning domain transformation through the LWLM encoder and DTI decoder, the trained LWLM can extract cross-domain consistent features that retain the essential channel information across domains.
Given a spatial-frequency domain channel representation $\bm{H}_s$, its corresponding angle-delay domain representation $T^{\text{DTI}}(\bm{H}_s)$ can be obtained by two-dimensional discrete Fourier transforms (2D-DFT). The transformation process can be expressed as
$   T^{\text{DTI}}(\bm{H}_s) = \bm{W}_{\theta}^{\mathsf{H}} \bm{H}_s \bm{W}_{\tau}^{*}$,
where $(\cdot)^{\mathsf{H}}$ denotes conjugate transpose and $(\cdot)^{*}$ denotes the complex conjugation. The unitary DFT matrices $\bm{W}_{\theta} \in \mathbb{C}^{N_{\text{ant}} \times N_{\text{ant}}}$ and $\bm{W}_{\tau} \in \mathbb{C}^{N_{\text{subc}} \times N_{\text{subc}}}$ map the spatial domain to the angle domain and the frequency domain to the delay domain, respectively. Each element of the DFT matrix can be expressed as
$[\bm{W}_\theta]_{i_1,i_2} = \frac{1}{\sqrt{N_{\text{ant}}}} \exp\!\left(-j 2\pi \frac{i_1 i_2}{N_{\text{ant}}} \right)$, and $[\bm{W}_\tau]_{i_1,i_2} = \frac{1}{\sqrt{N_{\text{subc}}}} \exp\!\left(-j 2\pi \frac{i_1 i_2}{N_{\text{subc}}} \right)
$.
To learn transformation-invariant features, the LWLM encoder first maps the input CFR into a latent representation: $\bm{o}^{\text{DTI}}_s = \mathcal{F}_{\bm{\Phi}}^{\text{LWLM}}(\bm{H}^{\text{DTI}}_s)$. Then, the DTI decoder, i.e. $\mathcal{F}^{\text{DTI}}_{\bm{\Theta}_{\text{DTI}}}(\cdot)$ with parameters $\bm{\Theta}_{\text{DTI}}$, decodes $\bm{o}^{\text{DTI}}_s$ to the angle-delay domain, i.e., 
$
    \hat{\bm{H}}^{\text{DTI}}_s = \mathcal{F}^{\text{DTI}}_{\bm{\Theta}_{\text{DTI}}}(\bm{o}^{\text{DTI}}_s)
$,
where $\hat{\bm{H}}^{\text{DTI}}_s$ denotes the output of the DTI decoder.

Following Theorem~\ref{theorem:1}, our target is to minimize the reconstruction loss. Considering the sparsity of the channel in the angle-delay domain, the reconstruction loss $\mathcal{L}_{\text{DTI}}$ is defined via angle-delay dissimilarity~\cite{stephan2024angle}:
\begin{align}
 \mathcal{L}_{\text{DTI}} = \frac{1}{N_{\text{bat}}} \!  \sum_{\bm{H}_s \in \mathcal{N}_{\text{bat}}}  \! [ 1  -
\frac{ \left| \langle T^{\text{DTI}}(\bm{H}_s), \hat{\bm{H}}^{\text{DTI}}_s \rangle \right| }
     { \| T^{\text{DTI}}(\bm{H}_s) \| \cdot \| \hat{\bm{H}}^{\text{DTI}}_s \|  } ],
\label{equ:LossDTI}
\end{align}
where $\langle \cdot, \cdot \rangle$ denotes the Frobenius inner product.

\subsection{PICL-based SSL}
\label{SubSec:PICL}

PICL leverages location invariance across different BSs or configurations using contrastive learning. By treating pairs of CSI observations from the same location as positive samples, the encoder learns to associate these views with similar latent representations. As shown in Theorem \ref{theorem:2}, contrastive-based PICL can enhance the expressiveness of the representation for task-specific semantics and make the semantics more suitable for localization tasks.

Given a CFR sample $\bm{H}_{s}$ and its corresponding positive sample $\bm{H}_{s^+}$, both of which are captured at the same location but under different BS deployments or configuration settings, these paired samples are used to guide the contrastive learning process in PICL\footnote{Note that for PICL, positive and negative samples are constructed based on the assumption that a UE can simultaneously obtain CSI measurements from multiple BSs or different configurations.}. This reflects a realistic acquisition process and does not rely on ground-truth labels during training. The LWLM encoder generates latent representations as follows:
\begin{align}
\bm{o}_{s}^{\text{PICL}} &= \mathcal{F}^{\text{LWLM}}_{\bm{\Phi}}(\bm{H}_{s}), \quad
\bm{o}_{s^+}^{\text{PICL}} = \mathcal{F}^{\text{LWLM}}_{\bm{\Phi}}(\bm{H}_{s^+}).
\end{align}
Latent representations $\bm{o}_{s}^{\text{PICL}}$ and $\bm{o}_{s^+}^{\text{PICL}}$ are then processed by PICL decoder $\mathcal{F}_{\bm{\Theta}_{\text{PICL}}}^{\text{PICL}}(\cdot)$, incorporating the associated BS configuration, to produce configuration-invariant representations:
\begin{align}
\! \! \! \tilde{\bm{o}}_{s}^{\text{PICL}} = \mathcal{F}_{\bm{\Theta}_{\text{PICL}}}^{\text{PICL}}(\bm{o}_{s}^{\text{PICL}}, \bm{c}_{s}), \
\tilde{\bm{o}}_{s^+}^{\text{PICL}} = \mathcal{F}_{\bm{\Theta}_{\text{PICL}}}^{\text{PICL}}(\bm{o}_{s^+}^{\text{PICL}}, \bm{c}_{s^+}),
\end{align}
where $\bm{c}_s = [\bm{p}^{\text{bs}}_s, B{\text{bw},s}$ includes the BS location and bandwidth configuration\footnote{In this paper, we consider only bandwidth variation, but this formulation can be easily extended to incorporate other configuration parameters such as frequency band, transmit power, etc.}.

During training, as described in Theorem~\ref{theorem:2}, we employ the NT-Xent loss \cite{chen2020simple}, which is a special form of the InfoNCE loss \cite{oord2018representation}. Specifically, two mini-batches of channel samples, $\mathcal{N}_1^{\text{bat}}$ and $\mathcal{N}_2^{\text{bat}}$, are sampled from pretraining datasets $\mathcal{D}^{\text{pre}}$, each containing $N_{\text{bat}}$ samples. The $s$-th sample in $\mathcal{N}_1^{\text{bat}}$ and the $s^+$-th sample in $\mathcal{N}_2^{\text{bat}}$ form a positive pair, while all others ($2N_{\text{bat}} - 2$) samples are treated as negatives. Let $\tilde{\mathcal{N}}^{\text{bat}} = \mathcal{N}_1^{\text{bat}} \cup \mathcal{N}_2^{\text{bat}}$ denote the full batch. The PICL loss is defined as
\begin{align}
\mathcal{L}_{\text{PICL}} = \frac{1}{2N_{\text{bat}}} \sum_{s \in \tilde{\mathcal{N}}^{\text{bat}}} \sum_{s^+ \in \tilde{\mathcal{N}}^{\text{bat}}} \mathbbm{1}_{s,s^+} \ell_{s,s^+},
\label{equ:LossPICL1}
\end{align}
where $\mathbbm{1}_{s,s^+} = 1$ if $s$ and $s^+$ form a positive pair, and $\mathbbm{1}_{s,s^+} = 0$ otherwise. For each positive pair $(s, s^+)$, the contrastive loss $\ell_{s,s^+}$ is given by
\begin{align}
\ell_{s,s^+} = -\log \frac{
\exp\left( \text{sim}(\tilde{\bm{o}}_{s}^{\text{PICL}}, \tilde{\bm{o}}_{s^+}^{\text{PICL}})/\tau \right)
}{
\sum\limits_{\substack{s^\star \in \tilde{\mathcal{N}}^{\text{bat}}, s^\star \neq s}} \exp\left( \text{sim}(\tilde{\bm{o}}_{s}^{\text{PICL}}, \tilde{\bm{o}}_{s^\star}^{\text{PICL}})/\tau \right)
}.
\label{equ:LossPICL2}
\end{align}
where $\text{sim}(\bm{a}, \bm{b}) = \frac{\bm{a}^\top \bm{b}}{\|\bm{a}\|\|\bm{b}\|}$ denotes cosine similarity and $\tau$ is a temperature hyperparameter.

\subsection{LWLM Architecture and Training Details}
\label{Subsec:Pretrain}

In this subsection, we provide a detailed description of the implementation and pretraining procedure of the proposed LWLM. Specifically, we describe the architecture of the encoder and SSL-specific decoders.

\subsubsection{Model Input}

As mentioned above, the input to the model is the complex CFR matrix $\bm{H}_s$. To facilitate neural network processing, we decompose it into its magnitude and phase components, denoted as $\overline{\bm{H}}_s$, and rewrite it as
\begin{align}
\overline{\bm{H}}_s = \left[|\bm{H}_s|, \angle \bm{H}_s\right] \in \mathbb{R}^{2 \times N_{\text{ant}} \times N_{\text{subc}}}. \label{euq:input}
\end{align}
The input $\overline{\bm{H}}_s$ is a 3D tensor of shape $(2, N_{\text{ant}}, N_{\text{subc}})$, with the first dimension corresponding to the amplitude and phase, respectively. Although $\overline{\bm{H}}_s$ represents a specific format of the channel input, for notational consistency throughout the paper, we will still use $\bm{H}_s$ to refer to the general representation of the input channel data in all subsequent discussions.

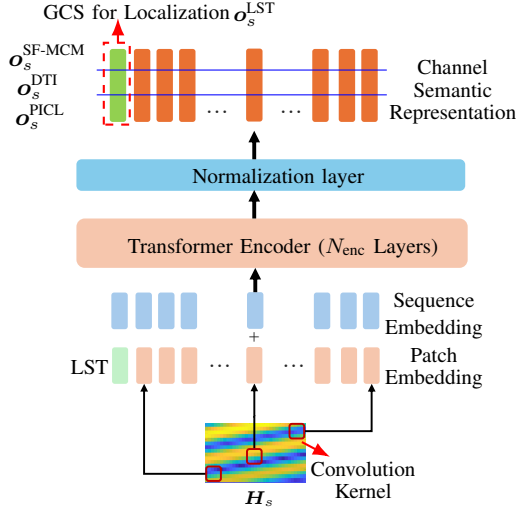
\begin{figure}[t]
    \vspace{2mm}
    \centering
    \begin{tikzpicture}
    \node (image) [anchor=south west]{\includegraphics[width=0.68\linewidth]{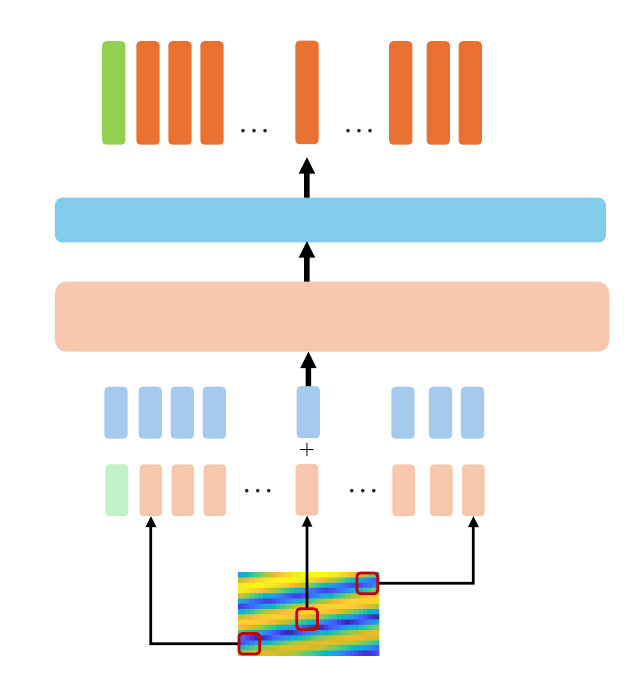}};
    \gettikzxy{(image.north east)}{\ix}{\iy};
    
    \node at (0.04*\ix,0.95*\iy)[rotate=0,anchor=north]{\small {$\bm{o}^{\text{SF-MCM}}_s$}};
    \node at (0.02*\ix,0.89*\iy)[rotate=0,anchor=north]{\small {$\bm{o}^{\text{DTI}}_s$}};
    \node at (0.025*\ix,0.83*\iy)[rotate=0,anchor=north]{\small {$\bm{o}^{\text{PICL}}_s$}};
    \draw[blue] (0.145*\ix, 0.83*\iy) -- (0.76*\ix, 0.83*\iy);
    \draw[blue] (0.145*\ix, 0.88*\iy) -- (0.76*\ix, 0.88*\iy);
    \draw[red, dashed, thick] (0.16*\ix, 0.77*\iy) rectangle (0.215*\ix, 0.935*\iy);
    \draw[red, ->, thick, >=Latex] (0.19*\ix, 0.935*\iy) -- (0.19*\ix, 0.985*\iy);

    \node at (0.23*\ix,1.03*\iy)[rotate=0,anchor=north]{\footnotesize \shortstack{GCS for Localization}};
    \node at (0.48*\ix,1.035*\iy)[rotate=0,anchor=north]{\footnotesize \shortstack{$\bm{o}_s^{\text{LST}}$}};
    \node at (0.90*\ix,0.92*\iy)[rotate=0,anchor=north]{\footnotesize  \shortstack{Channel \\ Semantic \\ Representation}};
    \node at (0.53*\ix,0.70*\iy)[rotate=0,anchor=north]{\footnotesize Normalization layer};
    \node at (0.54*\ix,0.56*\iy)[rotate=0,anchor=north]{\footnotesize Transformer Encoder ($N_{\text{enc}}$ Layers)};
    \node at (0.86*\ix,0.45*\iy)[rotate=0,anchor=north]{\footnotesize \shortstack{Sequence \\ Embedding}};
    \node at (0.86*\ix,0.34*\iy)[rotate=0,anchor=north]{\footnotesize \shortstack{Patch \\ Embedding}};
    \node at (0.71*\ix,0.11*\iy)[rotate=0,anchor=north]{\footnotesize \shortstack{Convolution\\Kernel}};
    \node at (0.49*\ix,0.05*\iy)[rotate=0,anchor=north]{\scriptsize \shortstack{${\bm{H}}_s$}};

    \draw[red, ->, thick, >=Latex] (0.58*\ix, 0.13*\iy) -- (0.64*\ix, 0.10*\iy);

    \node at (0.13*\ix,0.315*\iy)[rotate=0,anchor=north]{\footnotesize \shortstack{LST}};

    \end{tikzpicture}
    \vspace{-1mm}
    \caption{Schematic diagram of LWLM encoder architecture.}
    
    \label{fig:LWLM-Encoder}
\end{figure}

\subsubsection{Channel Embedding}

As shown in Fig.~\ref{fig:LWLM-Encoder}, before feeding into the transformer encoder, we perform channel tokenization to convert the 3D channel input ${\bm{H}}_s$ into a sequence of embeddings. Inspired by the tokens-to-token vision transformer (T2T-ViT) framework \cite{yuan2021tokens}, we apply a 2D CNN layer with $N_{\text{embed}}$ convolution kernels of size $K_{\text{cnn}} \times K_{\text{cnn}}$ and stride $S_{\text{cnn}}$ along both dimensions. This operation produces $N_{\text{patch}}$ tokens, each represented by an $N_{\text{embed}}$-dimensional vector. The total number of patches $N_{\text{patch}}$ is computed as
$
N_{\text{patch}} = \left\lfloor \frac{N_{\text{ant}} - K_{\text{cnn}}}{S_{\text{cnn}}} + 1 \right\rfloor \times \left\lfloor \frac{N_{\text{subc}} - K_{\text{cnn}}}{S_{\text{cnn}}} + 1 \right\rfloor.
$

Let $\bm{x}^{\text{embed}}_{s,u}$ denote the $u$-th token for sample $s$. To guide the model in learning global channel semantics (GCSs) for localization, inspired by BERT’s [CLS] token \cite{devlin2019bert}, we introduce a learnable embedding called the localization semantic token (LST). LST is denoted by $\bm{x}^{\text{embed}}_{l,0}$, and is placed at the beginning of the token sequence. As a trainable embedding, LST $\bm{x}^{\text{embed}}_{l,0}$ has no actual semantics before being input into the transformer encoder. This means that LST will not be biased towards its own local semantics like other tokens. After passing through multiple layers of transformers, it gradually integrates information from all tokens via the multi-head attention mechanism, thereby aggregating GCSs.
Additionally, since transformers are permutation-invariant, we incorporate sequence embeddings\footnote{In the context of transformers, this is commonly referred to as position embeddings \cite{vaswani2017attention}, indicating the sequential order of each patch. However, to avoid confusion with the position information in localization tasks, we term this as sequence embeddings.} to embed sequence-specific information. Consequently, the embedding sequence input $\bm{X}^{\text{embed}}_{s} \in \mathbb{R}^{(N_{\text{patch}}+1) \times N_{\text{embed}}}$ to the transformer is defined as
\begin{align}
\bm{X}^{\text{embed}}_{s} = [\bm{x}^{\text{embed}}_{l,0}, \bm{x}^{\text{embed}}_{l,1}, \dots, \bm{x}^{\text{embed}}_{l,N_{\text{patch}}}] + \bm{E}^{\text{seq}},
\label{equ:ChannelEmbed}
\end{align}
where $\bm{E}^{\text{seq}} \in \mathbb{R}^{(N_{\text{patch}} + 1) \times N_{\text{embed}}}$ is the sequence embedding matrix \cite{vaswani2017attention}, with entries defined as
\begin{align}
[\bm{E}^{\text{seq}}]_{seq, 2i} &= \sin\left(\frac{seq}{10000^{2i/N_{\text{embed}}}}\right), \\
[\bm{E}^{\text{seq}}]_{seq, 2i+1} &= \cos\left(\frac{seq}{10000^{2i/N_{\text{embed}}}}\right),
\end{align}
for sequence index $seq \in \{0, 1, \dots, N_{\text{patch}}\}$ and embedding dimension index $i$. This tokenization strategy transforms the 3D channel structure into a sequential format, enabling the transformer to capture both local and global dependencies across antennas and subcarriers.

\subsubsection{LWLM Encoder Architecture}

As illustrated in Fig.~\ref{fig:LWLM-Encoder}, the LWLM encoder $\mathcal{F}^{\text{LWLM}}_{\bm{\Phi}}(\cdot)$ consists of $N_{\text{enc}}$ transformer encoder layers followed by a normalization layer. Each transformer encoder layer includes multi-head self-attention modules and feed-forward neural networks to model long-range dependencies and semantic interactions across tokenized representations. Details of the transformer encoder layer can be found in \cite{vaswani2017attention}. The final normalization layer helps stabilize feature distributions and facilitates efficient training convergence.

The transformer encoder maps each $N_{\text{embed}}$-dimensional token to a semantic representation of dimension $N_{\text{latent}}$, resulting in an output matrix $\bm{o}_s \in \mathbb{R}^{(N_{\text{patch}}+1) \times N_{\text{latent}}}$.
We further partition the latent output $\bm{o}_s$ into three distinct parts to support the individual SSL objectives described in Sections~\ref{SubSec:SFMCM}, \ref{SubSec:DTI}, and \ref{SubSec:PICL}. The entire channel representation can be expressed as $
\bm{o}_s = [\bm{o}_s^{\text{SF-MCM}}, \bm{o}_s^{\text{DTI}}, \bm{o}_s^{\text{PICL}}],
$
where $\bm{o}_s^{\text{SF-MCM}} \in \mathbb{R}^{(N_\text{patch}+1) \times N_{\text{SF-MCM}}}$, $\bm{o}_s^{\text{DTI}} \in \mathbb{R}^{(N_\text{patch}+1) \times N_{\text{DTI}}}$, and $\bm{o}_s^{\text{PICL}} \in \mathbb{R}^{(N_{\text{patch}}+1) \times N_{\text{PICL}}}$ are the feature segments sent to their respective decoders. Therefore, the total dimension of $N_{\text{latent}}$ is given by
$
N_{\text{SF-MCM}} + N_{\text{DTI}} + N_{\text{PICL}} = N_{\text{latent}}.
$

In particular, the first row of $\bm{o}_s$, corresponding to the LST, is denoted by $\bm{o}_s^{\text{LST}}$, which serves as the GCS information. This vector captures high-level, location-relevant features and is used as input for downstream localization decoders. Detailed use of $\bm{o}_s^{\text{LST}}$ for localization tasks will be introduced in Sec.~\ref{sec:downstream_task}.

\subsubsection{Pretraining Decoder Architecture}

\begin{figure}[t]
    \centering
    \begin{tikzpicture}
    \node (image) [anchor=south west]{\includegraphics[width=0.95\linewidth]{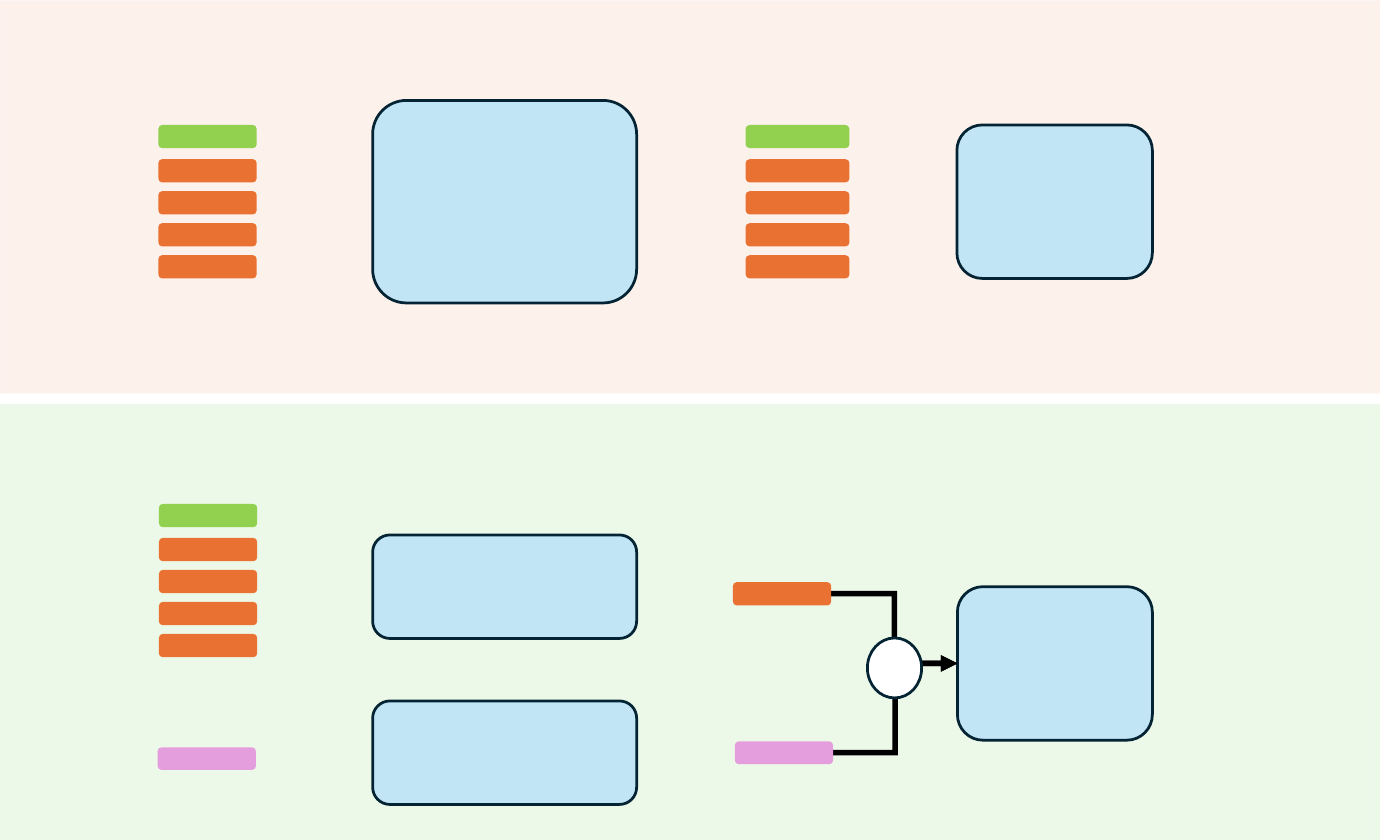}};
    \gettikzxy{(image.north east)}{\ix}{\iy};

    \node at (0.52*\ix, 0.98*\iy)[rotate=0,anchor=north]{\small \shortstack{SF-MCM/DTI Decoder}};
    \node at (0.17*\ix,0.925*\iy)[rotate=0,anchor=north]{{\footnotesize $\bm{o}^{\text{SF-MCM}}_s$/$\bm{o}^{\text{DTI}}_s$}};
    \node at (0.155*\ix,0.67*\iy)[rotate=0,anchor=north]{{\scriptsize \shortstack{$(N_{\text{bat}}, N_{\text{patch+1}},N_{\text{SF-MCM}})$ \\ $(N_{\text{bat}}, N_{\text{patch+1}},N_{\text{DTI}})$}}};
    \node at (0.37*\ix,0.84*\iy)[rotate=0,anchor=north]{{\footnotesize \shortstack{ Transformer \\ Decoder \\ ($N_{\text{dec}}$ Layers)}}};
    \node at (0.58*\ix,0.65*\iy)[rotate=0,anchor=north]{{\scriptsize $(N_{\text{bat}}, N_{\text{patch}},N_{\text{embed}})$ }};

    \node at (0.9*\ix,0.84*\iy)[rotate=0,anchor=north]{{\footnotesize $\hat{\bm{H}}^{\text{SF-MCM}}_s$}};
    \node at (0.87*\ix,0.75*\iy)[rotate=0,anchor=north]{{\footnotesize $\hat{\bm{H}}^{\text{DTI}}_s$}};
    \node at (0.86*\ix,0.67*\iy)[rotate=0,anchor=north]{{\scriptsize \shortstack{$(N_{\text{bat}}, 2, N_{\text{ant}},N_{\text{subc}})$}}};

    \draw[red, dashed, thick] (0.533*\ix, 0.65*\iy) rectangle (0.617*\ix, 0.805*\iy);

    \node at (0.51*\ix,0.52*\iy)[rotate=0,anchor=north]{\small PICL Decoder};
    \node at (0.18*\ix,0.50*\iy)[rotate=0,anchor=north]{\footnotesize {$\bm{o}^{\text{PICL}}_s$}};
    \node at (0.15*\ix,0.245*\iy)[rotate=0,anchor=north]{{\scriptsize $(N_{\text{bat}}, N_{\text{patch+1}},N_{\text{PICL}})$}};
    \node at (0.58*\ix,0.40*\iy)[rotate=0,anchor=north]{{\scriptsize $(N_{\text{bat}}, N_{\text{PICL}})$}};
    \node at (0.16*\ix,0.19*\iy)[rotate=0,anchor=north]{\footnotesize {$\bm{c}_s$}};
    \node at (0.16*\ix,0.11*\iy)[rotate=0,anchor=north]{{\scriptsize $(N_{\text{bat}}, 3)$}};
    \node at (0.58*\ix,0.11*\iy)[rotate=0,anchor=north]{{\scriptsize $(N_{\text{bat}}, N_{\text{PICL}})$}};
    \node at (0.37*\ix,0.355*\iy)[rotate=0,anchor=north]{{\footnotesize Mean Pooling}};
    \node at (0.37*\ix,0.16*\iy)[rotate=0,anchor=north]{{\footnotesize MLP Layer}};

    \node at (0.76*\ix,0.29*\iy)[rotate=0,anchor=north]{{\footnotesize \shortstack{MLP \\ Layer}}};
    \node at (0.76*\ix,0.815*\iy)[rotate=0,anchor=north]{{\footnotesize \shortstack{Fold \\ Layer}}};
    \node at (0.875*\ix,0.32*\iy)[rotate=0,anchor=north]{\footnotesize $\tilde{\bm{o}}_{s}^{\text{PICL}}$};
    \node at (0.905*\ix,0.22*\iy)[rotate=0,anchor=north]{{\scriptsize \shortstack{$(N_{\text{bat}},  \tilde{N}_{\text{PICL}})$}}};

    \node at (0.645*\ix,0.253*\iy)[rotate=0,anchor=north]{{+}};

    \draw[black, ->, thick, >=Latex] (0.20*\ix, 0.117*\iy) -- (0.27*\ix, 0.117*\iy);
    \draw[black, ->, thick, >=Latex] (0.465*\ix, 0.117*\iy) -- (0.525*\ix, 0.117*\iy);
    \draw[black, ->, thick, >=Latex] (0.20*\ix, 0.305*\iy) -- (0.27*\ix, 0.305*\iy);
    \draw[black, ->, thick, >=Latex] (0.465*\ix, 0.305*\iy) -- (0.525*\ix, 0.305*\iy);
    \draw[black, ->, thick, >=Latex] (0.83*\ix, 0.22*\iy) -- (0.95*\ix, 0.22*\iy);
    \draw[black, ->, thick, >=Latex] (0.20*\ix, 0.74*\iy) -- (0.27*\ix, 0.74*\iy);
    \draw[black, ->, thick, >=Latex] (0.465*\ix, 0.74*\iy) -- (0.537*\ix, 0.74*\iy);
    \draw[black, ->, thick, >=Latex] (0.615*\ix, 0.74*\iy) -- (0.685*\ix, 0.74*\iy);
    \draw[black, ->, thick, >=Latex] (0.83*\ix, 0.74*\iy) -- (0.95*\ix, 0.74*\iy);
    
    \end{tikzpicture}
    \vspace{-3mm}
    \caption{Schematic diagram of pretraining decoder architecture.}
    \label{fig:PretrainDecoder}
\end{figure}
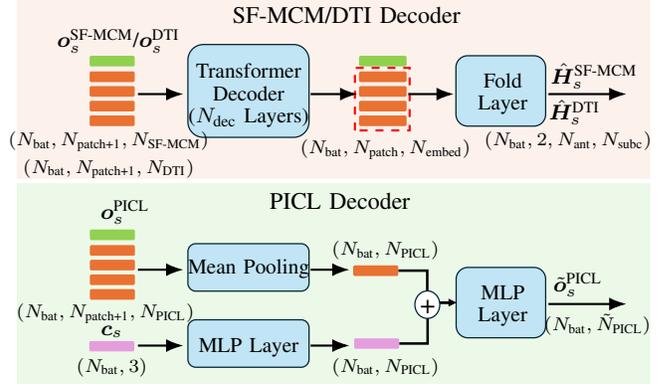

Fig.~\ref{fig:PretrainDecoder} illustrates the pretraining decoder architectures. The SF-MCM and DTI decoders have identical transformer-decoder-based structures with different parameters. Details of the transformer decoder can be found in \cite{vaswani2017attention}. Given semantic representations $\bm{o}_s^{\text{SF-MCM}}$ or $\bm{o}_s^{\text{DTI}}$, the transformer decoder with $N_{\text{dec}}$ layers reconstructs the channel representation. After decoding, The LST token is discarded, and the remaining tokens of shape $(N_{\text{bat}}, N_{\text{patch}}, N_{\text{embed}})$ are folded \cite{yuan2021tokens} to match the input dimensions. The final output is a 4-dimensional tensor $\hat{\bm{H}}_s^{\text{SF-MCM}}$ or $\hat{\bm{H}}_s^{\text{DTI}}$, as described in Sec. \ref{SubSec:SFMCM} and \ref{SubSec:DTI}. The output tensor has shape $(N_{\text{bat}}, 2, N_{\text{ant}}, N_{\text{subc}})$, where the second dimension represents the real and imaginary parts of the channel.

For the PICL decoder, the semantic representation $\bm{o}_s^{\text{PICL}}$ is first processed by a mean pooling operation over the patch dimension, producing a pooled channel representation of shape $(N_{\text{bat}}, N_{\text{PICL}})$. Simultaneously, the BS configuration vector $\bm{c}_s$ is passed through a single-layer MLP to generate a configuration embedding of the same shape as the pooled channel representation. These two embeddings are then combined via element-wise addition. The resulting vector is fed into another single-layer MLP, which outputs a final location-dependent embedding $\tilde{\bm{o}}_s^{\text{PICL}} \in \mathbb{R}^{N_{\text{bat}} \times \tilde{N}_{\text{PICL}}}$, where $\tilde{N}_{\text{PICL}}$ is an output dimension for downstream contrastive learning objectives.

\subsubsection{Pretraining Process}

\begin{algorithm}[tbp]
\caption{Self-Supervised Pretraining of LWLM}
\label{alg:pretrain}
\Input{$\mathcal{D}^{\text{pre}}$, $N_{\text{bat}}$, $\tau$, $\alpha_{\text{SF-MCM}}$, $\alpha_{\text{DTI}}$, $\alpha_{\text{PICL}}$}
\Output{$\bm{\Phi}$, $\bm{\Theta}_{\text{SF-MCM}}, \bm{\Theta}_{\text{DTI}}, \bm{\Theta}_{\text{PICL}}$}

\ForEach{training step}{
    Sample a batch $\mathcal{N}_1^{\text{bat}} \subseteq \mathcal{D}^{\text{pre}}$\;
    Sample another batch $\mathcal{N}_2^{\text{bat}} \subseteq \mathcal{D}^{\text{pre}}$ based on $\mathcal{N}_1^{\text{bat}}$\;

    // {\textbf{SF-MCM Task}}
    
    \ForEach{${\bm{H}}_s \in \mathcal{N}_1^{\text{bat}}$}{
        Apply random masking to obtain ${T}^{\text{SF-MCM}}(\bm{H}_s)$ using  \eqref{equ:mask}\;
        Generate $\bm{o}_s$ using $\mathcal{F}^{\text{LWLM}}_{\bm{\Phi}}(\cdot)$ and obtain $\bm{o}_s^{\text{SF-MCM}}$\;
        Reconstruct $\hat{\bm{H}}_s$ using $\mathcal{F}^{\text{SF-MCM}}_{\bm{\Theta}_{\text{SF-MCM}}}(\cdot)$\;
    }
    Compute SF-MCM loss $\mathcal{L}_{\text{SF-MCM}}$ using  \eqref{equ:LossSFMCM}\;

    // {\textbf{DTI Task}}
    
    \ForEach{${\bm{H}}_s \in \mathcal{N}_1^{\text{bat}}$}{
        Generate $\bm{o}_s$ using $\mathcal{F}^{\text{LWLM}}_{\bm{\Phi}}(\cdot)$ and obtain $\bm{o}_s^{\text{DTI}}$\;
        Reconstruct $\hat{\bm{H}}_s^{\text{DTI}}$ using $\mathcal{F}^{\text{DTI}}_{\bm{\Theta}_{\text{DTI}}}(\cdot)$\;
    }
    Compute DTI loss $\mathcal{L}_{\text{DTI}}$ using  \eqref{equ:LossDTI}\;

    // {\textbf{PICL Task}}
    
    \ForEach{${\bm{H}}_s \in \mathcal{N}_1^{\text{bat}} \cup \mathcal{N}_2^{\text{bat}}$}{
        Generate $\bm{o}_s$ using $\mathcal{F}^{\text{LWLM}}_{\bm{\Phi}}(\cdot)$ and obtain $\bm{o}_s^{\text{PICL}}$\;
        Obtain latent representation $\tilde{\bm{o}}^{\text{PICL}}_s$ using $\mathcal{F}^{\text{PICL}}_{\bm{\Theta}_{\text{PICL}}}(\cdot)$\;
    }
    Compute PICL loss $\mathcal{L}_{\text{PICL}}$ using  \eqref{equ:LossPICL1} and \eqref{equ:LossPICL2}\;

    Compute total loss $\mathcal{L}_{\text{pretrain}}$\;
    Update parameters $\bm{\Phi}$, $\bm{\Theta}_{\text{SF-MCM}}, \bm{\Theta}_{\text{DTI}}, \bm{\Theta}_{\text{PICL}}$ by minimizing $\mathcal{L}_{\text{pretrain}}$\;
}
\end{algorithm}

The pretraining procedure is summarized in Algorithm~\ref{alg:pretrain}, which outlines the detailed steps for optimizing LWLM under the proposed SSL method. First, a batch of channel samples $\mathcal{N}_1^{\text{bat}}$ is drawn from the unlabeled dataset $\mathcal{D}^{\text{pre}}$. Based on $\mathcal{N}_1^{\text{bat}}$, a second batch $\mathcal{N}_2^{\text{bat}}$ is constructed, where each sample in $\mathcal{N}_2^{\text{bat}}$ serves as a positive counterpart for the corresponding sample in $\mathcal{N}_1^{\text{bat}}$, as discussed in Sec.~\ref{SubSec:PICL}. Using $\mathcal{N}_1^{\text{bat}}$, we compute the SF-MCM and DTI losses $\mathcal{L}_{\text{SF-MCM}}$ and $\mathcal{L}_{\text{DTI}}$ as described in Eq.~\eqref{equ:LossSFMCM} and \eqref{equ:LossDTI}, respectively (see Sections~\ref{SubSec:SFMCM} and~\ref{SubSec:DTI}). For the PICL loss $\mathcal{L}_{\text{PICL}}$, both batches $\mathcal{N}_1^{\text{bat}}$ and $\mathcal{N}_2^{\text{bat}}$ are used jointly to compute the loss, as defined in Eq.~\eqref{equ:LossPICL1} and~\eqref{equ:LossPICL2}. For each task-specific loss, we first compute the full semantic representation $\bm{o}_s$ for every sample $s$, and extract the corresponding task-specific embeddings $\bm{o}_s^{\text{SF-MCM}}$, $\bm{o}_s^{\text{DTI}}$, and $\bm{o}_s^{\text{PICL}}$ as described in Sec~\ref{SubSec:SFMCM},~\ref{SubSec:DTI} and~\ref{SubSec:PICL}, and illustrated in Fig.~\ref{fig:LWLM-Encoder}. Finally, the total pretraining objective $\mathcal{L}_{\text{pretrain}}$ is defined as a weighted combination of the three pretraining losses:
\begin{align}
\mathcal{L}_{\text{pretrain}} = \alpha_{\text{SF-MCM}} \mathcal{L}_{\text{SF-MCM}} + \alpha_{\text{DTI}} \mathcal{L}_{\text{DTI}} + \alpha_{\text{PICL}} \mathcal{L}_{\text{PICL}},
\end{align}
where $\alpha_{\text{SF-MCM}},\ \alpha_{\text{DTI}},\  \alpha_{\text{PICL}}$ are predefined weighting coefficients that balance the contribution of each loss component.

\section{Downstream Localization-Relevant Tasks}
\label{sec:downstream_task}

To validate the effectiveness and generalization capability of the proposed LWLM, we evaluate its performance on four localization-related downstream tasks: ToA estimation, AoA estimation, single-BS localization, and multi-BS localization. Each task employs a lightweight task-specific decoder fine-tuned using a small amount of labeled data, while leveraging the pretrained LWLM encoder as the feature extractor. As mentioned above, we fine-tune both the encoder and decoder.

\subsection{ToA Estimation}

ToA estimation aims to predict the first-path propagation delay $\tau_{s,1}$ from the BS to the UE for channel $\bm{H}_s$ in LOS scenarios, which is often used to infer the radial distance between them. We first extract the GCS $\bm{o}_s^{\text{LST}}$ using the pretrained LWLM encoder $\mathcal{F}^{\text{LWLM}}_{\bm{\Phi}}(\cdot)$, as introduced in Sec.~\ref{Subsec:Pretrain}. The ToA is then predicted using a task-specific decoder $\mathcal{F}_{\bm{\Theta}_{\text{ToA}}}^{\text{d-task}}$ that takes both the semantic representation $\bm{o}_s^{\text{LST}} = \mathcal{F}^{\text{LWLM}}_{\bm{\Phi}}(\bm{H}_s)$ and the BS configuration $\bm{c}_s$ as input:
\begin{align}
    \hat{\tau}_{s,1} = \mathcal{F}_{\bm{\Theta}_{\text{ToA}}}^{\text{d-task}}(\bm{o}_s^{\text{LST}}, \bm{c}_s).
\end{align}

The decoder is implemented by MLPs as shown in Fig. \ref{fig:SBdecoder}. The BS configuration $\bm{c}_s$ uses a single-layer MLP embedding and adds it to $\bm{o}_s^{\text{LST}}$, which is then estimated by a two-layer regression MLP. The training objective is to minimize the mean absolute error (MAE):
\begin{align}
    \mathcal{L}_{\text{ToA}} = \frac{1}{N_{\text{bat}}} \sum_{s \in \mathcal{N}_{\text{bat}}} \left| \hat{\tau}_{s,1} - \tau_{s,1} \right|.
    \label{equ:Losstoa}
\end{align}

\subsection{AoA Estimation}

\begin{figure}[t]
    \centering
    \begin{tikzpicture}
    \node (image) [anchor=south west]{\includegraphics[width=0.94\linewidth]{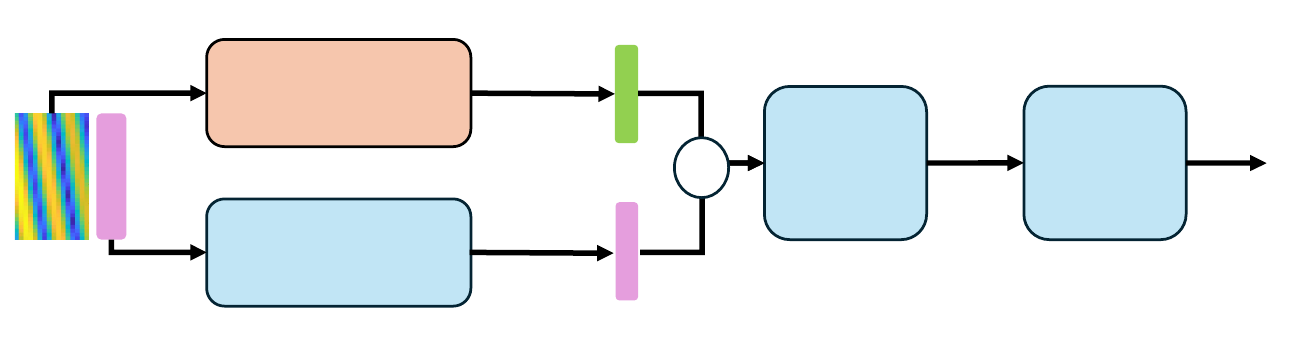}};
    \gettikzxy{(image.north east)}{\ix}{\iy};

    \node at (0.02*\ix,0.873*\iy)[rotate=0,anchor=north]{\small \shortstack{$\bm{H}^{\text{BS}}_s$}};
    \node at (0.14*\ix,0.30*\iy)[rotate=0,anchor=north]{\small \shortstack{$\bm{c}_s$}};

    \node at (0.268*\ix,0.87*\iy)[rotate=0,anchor=north]{\small \shortstack{LWLM \\ Encoder}};

    \node at (0.43*\ix,0.91*\iy)[rotate=0,anchor=north]{\small \shortstack{$\bm{o}_s^{\text{LST}}$}};

    \node at (0.267*\ix,0.38*\iy)[rotate=0,anchor=north]{\small \shortstack{MLP Layer}};

    \node at (0.645*\ix,0.70*\iy)[rotate=0,anchor=north]{\small \shortstack{MLP \\ Layer}};
    \node at (0.84*\ix,0.70*\iy)[rotate=0,anchor=north]{\small \shortstack{MLP \\ Layer}};

    \node at (0.985*\ix,0.81*\iy)[rotate=0,anchor=north]{\small \shortstack{$\hat{\tau}_{s,1}$ \\ $\hat{\theta}_{s,1}$ \\ $\hat{\bm{p}}_{s}^{\text{ue}}$}};

    \node at (0.537*\ix,0.597*\iy)[rotate=0,anchor=north]{{+}};

    \end{tikzpicture}
    \vspace{-4mm}
    \caption{Schematic diagram of the encoder-decoder architecture for ToA estimation, AoA estimation, and single-BS localization. Blue boxes indicate the decoder components.}
    \vspace{-1mm}
    \label{fig:SBdecoder}
\end{figure}
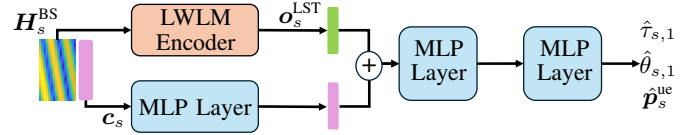

Similar to ToA estimation, AoA estimation aims to determine the azimuth angle $\theta_{s,1}$ of the first path from the UE to the BS in LOS scenarios. Based on channel $\bm{H}_s$, we use the LWLM encoder to extract the GCS $\bm{o}_s^{\text{LST}} = \mathcal{F}^{\text{LWLM}}_{\bm{\Phi}}(\bm{H}_s)$. The AoA is then estimated by a dedicated decoder $\mathcal{F}_{\bm{\Theta}_{\text{AoA}}}^{\text{d-task}}$:
\begin{align}
    \hat{\theta}_{s,1} = \mathcal{F}_{\bm{\Theta}_{\text{AoA}}}^{\text{d-task}}(\bm{o}_s^{\text{LST}}, \bm{c}_s).
\end{align}
The decoder architecture is similar to that of ToA, as shown in Fig. \ref{fig:SBdecoder}. The loss function is also based on MAE:
\begin{align}
    \mathcal{L}_{\text{AoA}} = \frac{1}{N_{\text{bat}}} \sum_{s \in \mathcal{N}_{\text{bat}}} \left| \hat{\theta}_{s,1} - \theta_{s,1} \right|.
    \label{equ:LossaAoA}
\end{align}

\subsection{Single-BS Localization}

Single-BS localization aims to estimate the UE position based solely on the CSI received from a single BS. Given a CFR sample $\bm{H}_s$, the LWLM encoder extracts the GCS representation $\bm{o}_s^{\text{LST}}$. A task-specific decoder $\mathcal{F}_{\bm{\Theta}_{\text{SB-Loc}}}^{\text{d-task}}$ then predicts the UE location $\hat{\bm{p}}_s^{\text{ue}}$:
\begin{align}
    \hat{\bm{p}}_s^{\text{ue}} = \mathcal{F}_{\bm{\Theta}_{\text{SB-Loc}}}^{\text{d-task}}(\bm{o}_s^{\text{LST}}, \bm{c}_s),
\end{align}

As illustrated in Fig.~\ref{fig:SBdecoder}, the decoder shares the same architecture as those used for ToA and AoA estimation. It consists of a configuration embedding module (a single-layer MLP) followed by a two-layer MLP with output dimension $d_p$. The configuration embedding is first combined with the semantic vector $\bm{o}_s^{\text{LST}}$ via element-wise addition, and the resulting feature is used for position regression. The training loss is the mean Euclidean distance:
\begin{align}
    \mathcal{L}_{\text{loc}} = \frac{1}{N_{\text{bat}}} \sum_{s \in \mathcal{N}_{\text{bat}}} \left\| \hat{\bm{p}}_s^{\text{ue}} - \bm{p}_s^{\text{ue}} \right\|_2.
    \label{equ:LossLoc}
\end{align}

\subsection{Multi-BS Localization}

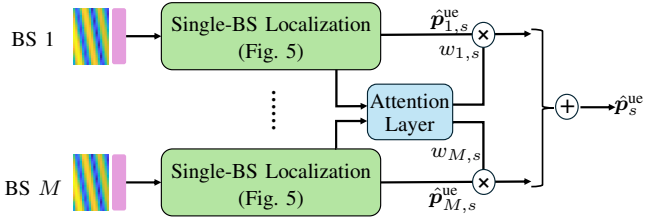
\begin{figure}[t]
    \centering
    \begin{tikzpicture}
    \node (image) [anchor=south west]{\includegraphics[width=0.95\linewidth]{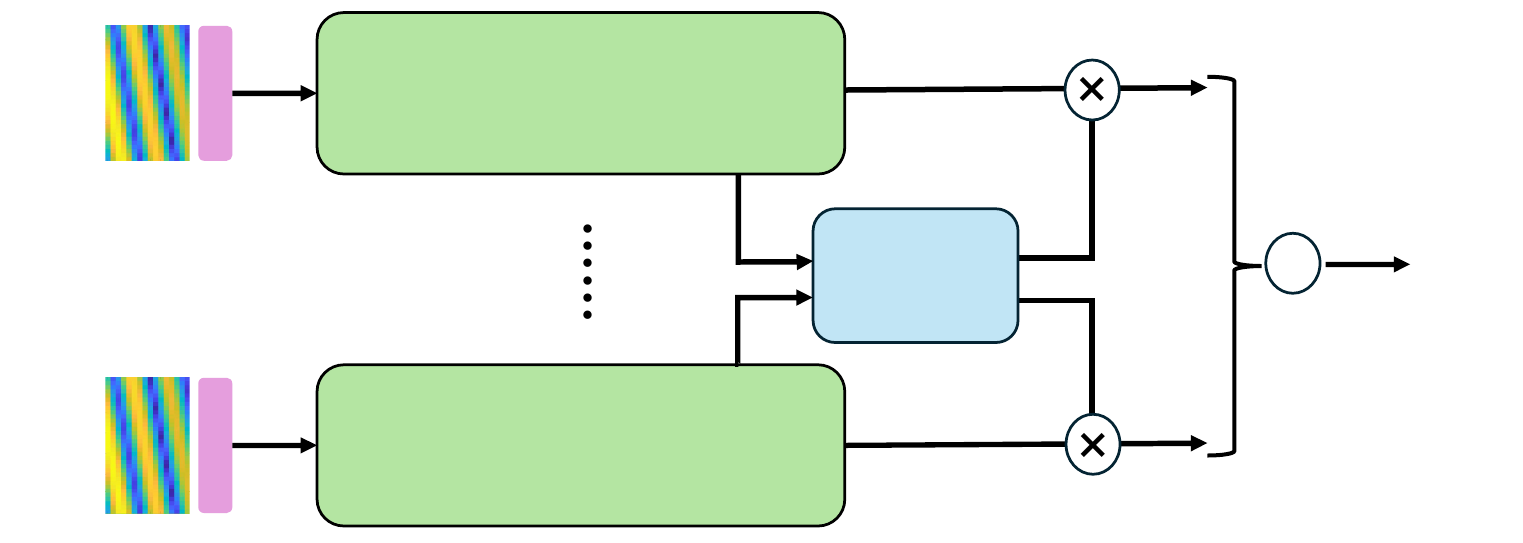}};
    \gettikzxy{(image.north east)}{\ix}{\iy};

    \node at (0.02*\ix,0.25*\iy)[rotate=0,anchor=north]{\footnotesize \shortstack{BS $M$}};
    \node at (0.385*\ix,0.32*\iy)[rotate=0,anchor=north]{\footnotesize \shortstack{Single-BS Localization \\
(Fig.~\ref{fig:SBdecoder})}};

    \node at (0.02*\ix,0.86*\iy)[rotate=0,anchor=north]{\footnotesize \shortstack{BS 1}};
    \node at (0.385*\ix,0.92*\iy)[rotate=0,anchor=north]{\footnotesize \shortstack{Single-BS Localization \\
(Fig.~\ref{fig:SBdecoder})}};

    \node at (0.598*\ix,0.605*\iy)[rotate=0,anchor=north]{\footnotesize \shortstack{Attention \\
Layer}};
    \node at (0.66*\ix,0.95*\iy)[rotate=0,anchor=north]{\footnotesize $\hat{\bm{p}}_{1,s}^{\text{ue}}$};
    \node at (0.675*\ix,0.80*\iy)[rotate=0,anchor=north]{\footnotesize $w_{1,s}$};

    \node at (0.665*\ix,0.22*\iy)[rotate=0,anchor=north]{\footnotesize $\hat{\bm{p}}_{M,s}^{\text{ue}}$};
    \node at (0.67*\ix,0.385*\iy)[rotate=0,anchor=north]{\footnotesize $w_{M,s}$};

    \node at (0.935*\ix,0.60*\iy)[rotate=0,anchor=north]{\footnotesize $\hat{\bm{p}}_{s}^{\text{ue}}$};

    \node at (0.84*\ix,0.58*\iy)[rotate=0,anchor=north]{{+}};

    \end{tikzpicture}
    \vspace{-3mm}
    \caption{Schematic diagram of multi-BS localization decoder architecture.}
    
    \label{fig:MB_Loc}
\end{figure}

Although single-BS localization performs well in certain cases, it often suffers in multipath-rich or NLoS scenarios. In contrast, multi-BS localization aggregates CSI from multiple BSs to jointly estimate the UE position, thereby improving robustness and accuracy.

Let $M$ denote the number of participating BSs. For a fixed UE location, the set of CFR samples collected from $M$ BSs is denoted as $\{\bm{H}_{m,s}\}_{m=1}^{M}$. Each channel sample is independently encoded by the LWLM encoder to obtain the corresponding GCS representation $\bm{o}_{m,s}^{\text{LST}}$.

To maintain scalability with varying numbers of BSs, we extend the single-BS decoder design, as shown in Fig.~\ref{fig:SBdecoder}. Specifically, each BS independently estimates a candidate UE position $\hat{\bm{p}}_{m,s}^{\text{ue}}$ using a dedicated single-BS decoder $\mathcal{F}_{\bm{\Theta}_{\text{SB-Loc},m}}^{\text{d-task}}$, which operates on the semantic vector $\bm{o}_{m,s}^{\text{LST}}$ and its corresponding BS configuration $\bm{c}_{m,s}$:
\begin{align}
    \hat{\bm{p}}_{m,s}^{\text{ue}} = \mathcal{F}_{\bm{\Theta}_{\text{SB-Loc},m}}^{\text{d-task}}(\bm{o}_{m,s}^{\text{LST}}, \bm{c}_{m,s}).
\end{align}
An attention-based fusion mechanism is subsequently applied to aggregate position estimates from all BSs. The final UE position is computed as a weighted sum of the individual predictions:
\begin{align}
    \hat{\bm{p}}_s^{\text{ue}} = \sum_{m=1}^{M} w_{m,s} \cdot \hat{\bm{p}}_{m,s}^{\text{ue}},
\end{align}
where $w_{m,s}$ denotes the attention weight associated with BS $m$.
The attention weight $w_{m,s}$ is derived from the feature vector $\bm{x}_{m,s}^{\text{SB}}$ extracted from the penultimate layer of the corresponding single-BS localization decoder. We employ a lightweight MLP as the attention layer, denoted as $\text{MLP}_{\text{attn}}(\cdot)$, to compute the attention logits:
\begin{align}
    w_{m,s} = \frac{\exp\left( \text{MLP}_{\text{attn}}(\bm{x}_{m,s}^{\text{SB}}) \right)}{\sum_{m^*=1}^{M} \exp\left( \text{MLP}_{\text{attn}}(\bm{x}_{m^*,s}^{\text{SB}}) \right)}.
    \label{equ：weight}
\end{align}
This attention mechanism enables the model to assign adaptive confidence weights to each BS, effectively leveraging the most reliable information sources in challenging environments.

Consequently, the overall multi-BS localization decoder $\mathcal{F}_{\bm{\Theta}_{\text{MB-Loc}}}$ is consists of two key components: (i) Single-BS decoders $\{ \mathcal{F}_{\bm{\Theta}_{\text{SB-Loc},m}}^{\text{d-task}} \}_{m=1}^{M}$ and (ii) an attention-based fusion module implemented via $\text{MLP}_{\text{attn}}(\cdot)$. During fine-tuning, we only need to train the attention layer once, even for different numbers of participating BSs. The output of the attention layer can be normalized by \eqref{equ：weight} to adapt to different numbers of BSs. In addition, during actual deployment, the localization results and weights of each BS can be transmitted to the fusion center or the user side for aggregation, effectively reducing the data transmission overhead. This modular design offers both flexibility and scalability, making it well-suited for deployment in dynamic wireless environments with varying numbers of BSs. Similar to the single-BS localization, we use Eq. \eqref{equ:LossLoc} as the loss function.

\section{Experiments}
\label{sec:Experiments}
\subsection{Experiment Settings}

\begin{figure}[t]
    \centering
    \begin{tikzpicture}
    \node (image) [anchor=south west]{\includegraphics[width=0.77\linewidth]{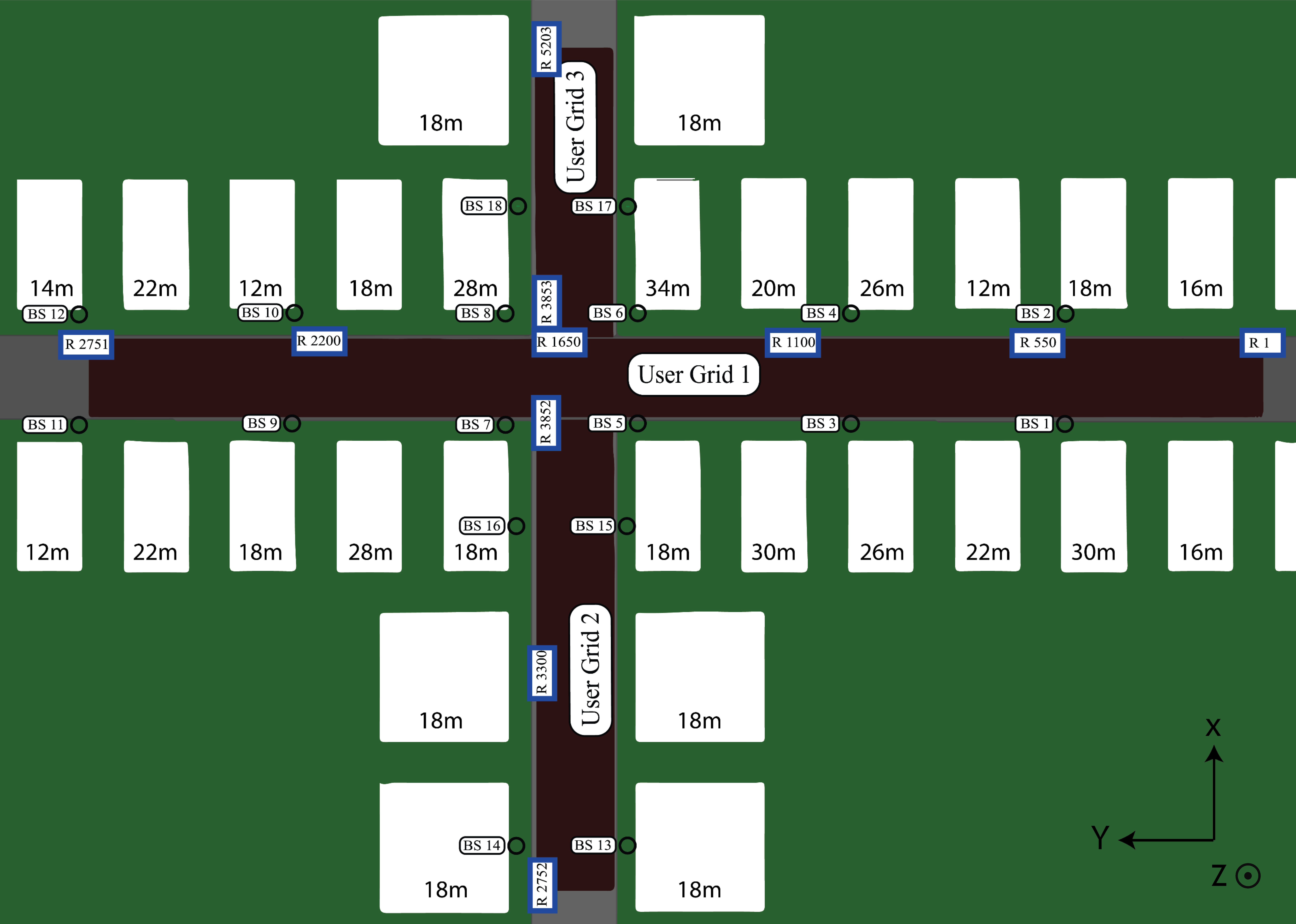}};
    \gettikzxy{(image.north east)}{\ix}{\iy};
    \draw[red, dashed, thick] (0.1*\ix, 0.55*\iy) rectangle (0.96*\ix, 0.63*\iy);
    \draw[red, thick] (0.60*\ix, 0.645*\iy) rectangle (0.66*\ix, 0.665*\iy);
    \draw[red, thick] (0.18*\ix, 0.645*\iy) rectangle (0.24*\ix, 0.665*\iy);
    \draw[red, thick] (0.18*\ix, 0.53*\iy) rectangle (0.24*\ix, 0.55*\iy);
    \draw[red, thick] (0.60*\ix, 0.53*\iy) rectangle (0.66*\ix, 0.55*\iy);

    \end{tikzpicture}
    \vspace{-1mm}
    \caption{Schematic diagram of channel simulation scenario O1. The red dashed box is the sampling area of the channel, and the red solid box represents the 4 BSs involved in channel collection.}
    \vspace{-3mm}
    \label{fig:environment}
\end{figure}
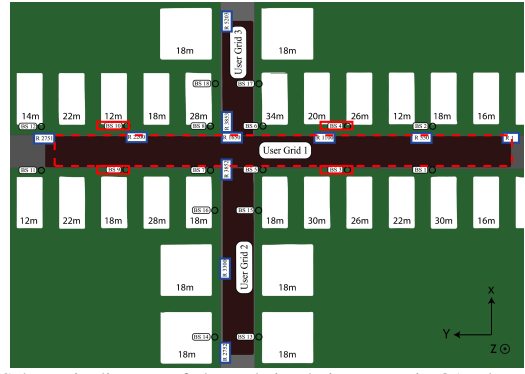

\begin{table}[tbp]
\centering
\caption{System and model parameters}
\label{tab:system_params}
\begin{tabular}{ll}
\toprule
\textbf{Parameter} & \textbf{Value} \\
\midrule
 Number of antennas/subcarriers ($N_{\text{ant}}$/$N_{\text{subc}}$) & 32/128 \\
 BS bandwidth ($B_{\text{bw}}$)  & 10/20/50 MHz \\
 2D CNN kernel size ($K_{\text{cnn}}$)  & 6 \\
 2D CNN stride ($S_{\text{cnn}}$) & 4 \\
 Number of patches ($N_{\text{patch}}$) & 256 \\
 Number of Transformer encoder layers ($N_{\text{enc}}$)  & 4 \\
 Number of Transformer decoder layers ($N_{\text{dec}}$)  & 2 \\
 Number of masked antennas/subcarriers ($\hat{N}_{\text{ant}}^{M}$/$\hat{N}_{\text{subc}}^{M}$) & 16/64\\
 Number of heads in Transformer encoder/decoder & 4 \\
 Embedding dimension ($N_{\text{embed}}$) & 512 \\
 Latent dimension ($N_{\text{latent}}$) & 256 \\
 Batch size ($N_{\text{bat}}$) & 32 \\
 SF-MCM latent dimension ${\bm{o}}_{s}^{\text{SF-MCM}}$ ($N_{\text{SF-MCM}}$) & 96 \\
 DTI latent dimension ${\bm{o}}_{s}^{\text{DTI}}$ ($N_{\text{DTI}}$)  & 96 \\
 PICL latent dimension ${\bm{o}}_{s}^{\text{PICL}}$ ($N_{\text{PICL}}$)  & 64 \\
 PICL latent dimension for $\tilde{\bm{o}}_{s}^{\text{PICL}}$ ($\tilde{N}_{\text{PICL}}$)   & 32 \\
 Pretraining epochs    & 200 \\
 Fine-tuning epochs    & 1000 \\
 Learning rate         & 0.0001 \\
 Loss function weight  ($\alpha_{\text{SF-MCM}},\ \alpha_{\text{DTI}},\  \alpha_{\text{PICL}}$)        & 10, 20, 1\\
 LWLM encoder parameters & 5.27 M \\
ToA/AoA/Single-BS localization decoder parameters & 0.07 M \\
\bottomrule
\end{tabular}
\end{table}

In this study, we use DeepMIMO \cite{Alkhateeb2019}, a widely adopted ray-tracing-based channel dataset, to evaluate our proposed wireless localization algorithms. Specifically, we select the O1 scenario \cite{Alkhateeb2019}, which models a typical outdoor urban environment with two intersecting streets, as shown in Fig.~\ref{fig:environment}.

For the self-supervised pretraining phase, we consider 4 BSs (i.e., BS 3, 4, 9, and 10). Each BS is equipped with a ULA comprising 32 antennas and utilizes 128 subcarriers. The channel bandwidths used during the pretraining phase are set to 10 MHz, 20 MHz, and 50 MHz. Channel data are collected from the spatial region highlighted by the red rectangle along the horizontal street shown in Fig. \ref{fig:environment}. For each BS configuration (combination of BS location and bandwidth setting), we collect channel measurements from 488,698 unique spatial locations, resulting in a comprehensive pretraining dataset containing $4 \times 3 \times 488,698 = 5,864,376$ channel samples. In the fine-tuning stage, unless otherwise specified, we randomly select channel samples from $10,000$ locations for training, channel samples from $1,000$ locations for validation, and an additional 10,000 locations for testing the performance of our model. System parameters and model parameters deployment details are shown in Table \ref{tab:system_params}. In the ToA estimation, AoA estimation, and single BS localization experiments, we take the performance of BS 3 as a representative case study to evaluate and compare the effectiveness of the proposed method.

We evaluate the performance of the following algorithms:
\begin{itemize}
\item \textbf{LWLM}: The proposed model trained with our hybrid self-supervised pretraining approach combining SF-MCM, DTI, and PICL.
\item \textbf{LWLM-SFMCM}: Only the SF-MCM-based pretraining strategy is used as an ablation experiment.
\item \textbf{LWLM-DTI}: Only the DTI-based pretraining strategy is used as an ablation experiment.
\item \textbf{LWLM-PICL}: Only the PICL-based pretraining strategy is used as an ablation experiment.
\item \textbf{LWLM-Base}: The proposed neural architecture trained directly on the localization task without any pretraining.
\item \textbf{CNN}: Use ResNet-34 \cite{wu2021learning}, with 21.28 M parameters, as the backbone network model architecture for localization.
\item \textbf{OMP}: A classical model-based localization algorithm employing OMP, which estimates angles and distances between the user and the BS. For multi-BS localization tasks, OMP estimates from multiple BSs are averaged to obtain the final location estimation.
\end{itemize}

\subsection{ToA Estimation Results}

\begin{figure}[t]
\centering
\includegraphics[scale=0.85]{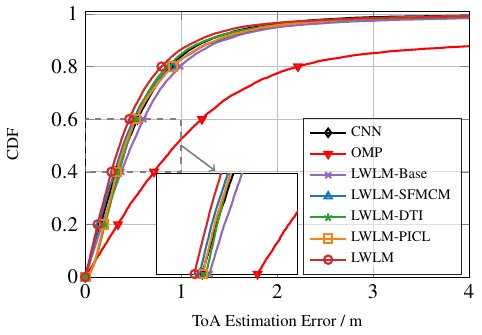}
\caption{CDF of ToA estimation error with 10,000 fine-tuning samples.}
\label{fig:ToA_CDF}
\end{figure}

Fig.~\ref{fig:ToA_CDF} presents the cumulative distribution function (CDF) of ToA estimation errors across different methods. Among the single-objective pretraining methods, LWLM-SFMCM, LWLM-DTI, and LWLM-PICL achieve 50\% ToA estimation errors of 0.39~m, 0.43~m, and 0.41~m, respectively. The LWLM-Base model without any pretraining yields a 50\% error of 0.48~m. Our proposed LWLM, which integrates all three SSL objectives, achieves the best performance with a 50\% ToA estimation error of only 0.36~m. This corresponds to a reduction of approximately 9.0\%--16.7\% in error compared to the single-objective variants, and a 26.0\% reduction relative to LWLM-Base. Furthermore, LWLM significantly outperforms both the CNN-based method and the model-based OMP algorithm, achieving performance improvements of 16.0\% and 62.1\%, respectively.

\begin{figure}[t]
\centering
\includegraphics[scale=0.85]{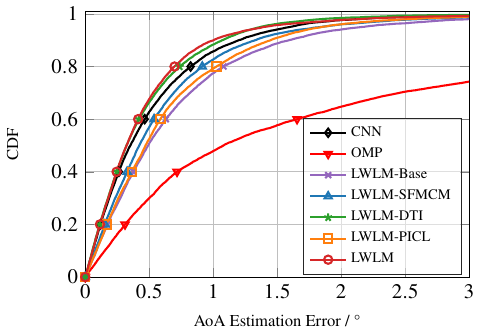}
\caption{CDF of AoA estimation error with 10,000 fine-tuning samples.}
\label{fig:AoA_CDF}
\end{figure}

\subsection{AoA Estimation Results}

Fig.~\ref{fig:AoA_CDF} compares the AoA estimation performance across different models in terms of the CDF of estimation errors. Among all methods, our proposed LWLM achieves the best result, with a 50\% AoA error of only 0.32°, indicating its strong capability in capturing angle-related semantic features from wireless channels. The LWLM-Base model without pretraining reaches a 50\% error of 0.49°, while the single-objective pretrained models, i.e., LWLM-SFMCM, LWLM-DTI, and LWLM-PICL, show 0.36°, 0.32°, and 0.47°, respectively. LWLM achieves a 34.7\% reduction in median AoA error compared to LWLM-Base. Furthermore, LWLM also outperforms CNN and OMP methods, providing 25.9\% and 75.8\% reductions in error, respectively.

\subsection{Single-BS Localization Results}

\begin{figure*}[t]
\centering
\captionsetup[subfigure]{font=small, skip=5pt}

\subfloat[\footnotesize CDF of localization error with 1 BS and 10,000 fine-tuning samples.\label{fig:SB_result_CDF}]{
    \includegraphics[width=0.79\columnwidth]{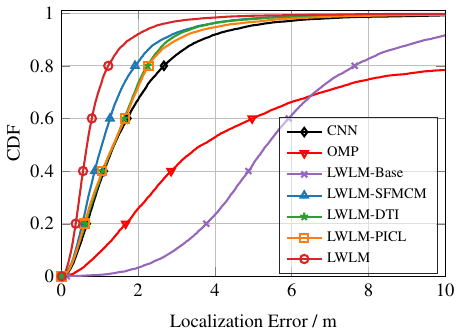}
    \vspace{-2mm}
}
\hspace{0.12\textwidth}
\subfloat[\footnotesize Average localization error results under different fine-tuning dataset size for single-BS localization.\label{fig:SB_result_dataset}]{
    \includegraphics[width=0.75\columnwidth]{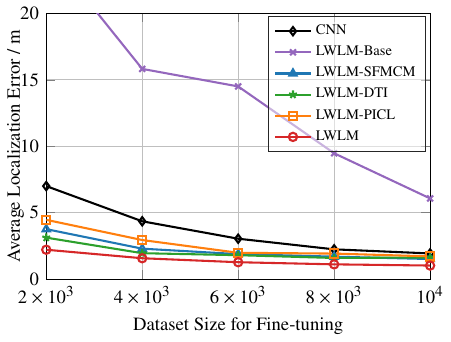}
    \vspace{-2mm}
}

\vspace{-4mm}

\subfloat[\footnotesize Average localization error results under different bandwidths for single-BS localization.\label{fig:SB_result_BW}]{
    \includegraphics[width=0.75\columnwidth]{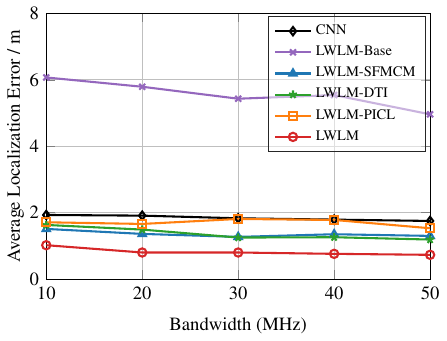}
    \vspace{-2mm}
}
\hspace{0.12\textwidth}
\subfloat[\footnotesize Average localization error results under different pilot settings for single-BS localization.\label{fig:SB_result_pilot}]{
    \includegraphics[width=0.75\columnwidth]{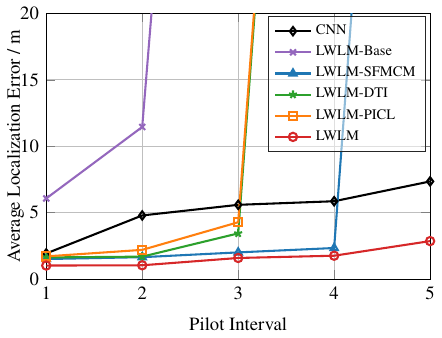}
    \vspace{-2mm}
}

\caption{Performance evaluation of single-BS localization tasks under different settings.}
\label{fig:SB_combined}
\vspace{-6mm}
\end{figure*}

Fig.~\ref{fig:SB_combined}\subref{fig:SB_result_CDF} presents the CDF curves of localization errors for BS 3 at a bandwidth of 10 MHz and with 10,000 fine-tuning samples. As illustrated in the figure, all three pretraining methods (i.e., LWLM-SFMCM, LWLM-DTI, and LWLM-PICL) individually achieve significantly lower localization errors compared to models without pretraining (i.e., LWLM-Base and CNN). Moreover, our proposed LWLM, which integrates all three self-supervised pretraining approaches, further enhances localization accuracy. Specifically, the proposed LWLM can achieve an average localization error of 0.67 m, which reduces the localization error by about 81.8\% compared with the model-based OMP algorithm. Additionally, LWLM achieves about 51.4\%–87.5\% improvement over LWLM-Base and CNN without pretraining. Even compared to the single-pretraining method, LWLM further reduces the localization error by 36.7\%–51.1\%, demonstrating the effectiveness and superiority of our proposed hybrid self-supervised approach in single-BS localization scenarios.

To evaluate generalization under limited supervision during the fine-tuning phase, we compare localization accuracy across different fine-tuning sample sizes, as shown in Fig.~\ref{fig:SB_combined}\subref{fig:SB_result_dataset}. We observe that the localization error consistently decreases as the amount of training data increases. However, when fine-tuning data is scarce, the advantages of self-supervised pretraining become particularly prominent. Notably, our proposed LWLM model, utilizing a hybrid pretraining strategy, achieves an average localization error of only 2.21 m with only 2,000 fine-tuning samples, significantly outperforming non-pretrained models and those baselines using a single pretraining method. These results confirm the strong generalization capability of LWLM in label-limited localization scenarios.

Fig.~\ref{fig:SB_combined}\subref{fig:SB_result_BW} illustrates the localization performance under varying bandwidth configurations. As the bandwidth increases, the temporal resolution of the channel improves, resulting in lower localization error across all models. Importantly, our pretraining dataset includes only 10~MHz, 20~MHz, and 50~MHz bandwidths, whereas the 30~MHz and 40~MHz configurations were excluded from pretraining. Despite this, LWLM maintains strong performance across all bandwidth settings. With 30MHz bandwidth, LWLM can achieve an average localization error of 0.80 m, which is 36.0-55.8\% lower than other methods based on a single pretraining method. This result demonstrates the robustness of our pretrained model and its strong transferability across unseen deployment conditions.

To simulate practical cellular systems such as LTE and 5G, where users typically obtain CSI only at pilot positions, we further evaluate localization performance under comb-type pilot configurations. Different from Fig.~\ref{fig:SB_combined}\subref{fig:SB_result_BW}, Fig.~\ref{fig:SB_combined}\subref{fig:SB_result_pilot} shows the performance of different pilot configurations with 10 MHz bandwidth. In these configurations, pilots are inserted every $N_{\text{pilot}}$ subcarriers or antennas in the spatial-frequency domain, while the other positions of subcarriers and antennas are filled with 0. We simulate pilot intervals $N_{\text{pilot}} = \{1, 2, 3, 4, 5\}$. Here, $N_{\text{pilot}} = 1$ indicates that pilot signals are present at all positions of the channel matrix. As the pilot interval increases, localization performance degrades due to the reduced density of channel information. When $N_{\text{pilot}} = 3$, the non-pretrained model LWLM-Base suffers substantial performance loss. Models with DTI or PICL pretraining maintain reasonable performance up to $N_{\text{pilot}} = 3$, but degrade rapidly beyond that. In contrast, the SF-MCM pretraining strategy, which is based on masked modeling aligned with the sparse nature of pilot configurations, consistently outperforms other baselines. However, even SF-MCM exhibits unacceptable localization error when $N_{\text{pilot}} \ge 4$. Most notably, our proposed LWLM method demonstrates the highest robustness and adaptability. Even with $N_{\text{pilot}} = 5$, LWLM achieves an average localization error of only 2.86 m, significantly outperforming all baselines. This underscores the strong synergy between masked channel modeling and invariant feature learning in addressing sparse pilot and measurement challenges in real-world cellular localization scenarios.

\subsection{Multi-BS Localization Results}

\begin{figure}[t]
\centering
\captionsetup[subfigure]{font=small, skip=5pt}

\subfloat[\footnotesize CDF of localization error with 4 BSs and 10,000 fine-tuning samples.\label{fig:MB_result_CDF}]{
    \includegraphics[width=0.40\textwidth]{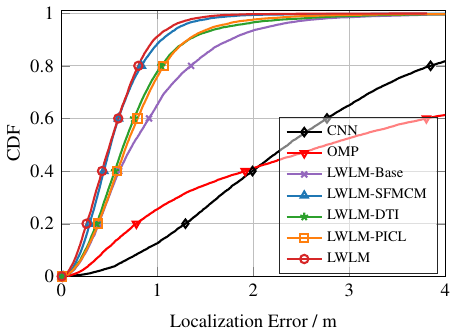}
    \vspace{-3mm}
}

\vspace{-2mm}

\subfloat[\footnotesize Average localization error results under different numbers of BSs.\label{fig:MB_basestations}]{
    \includegraphics[width=0.37\textwidth]{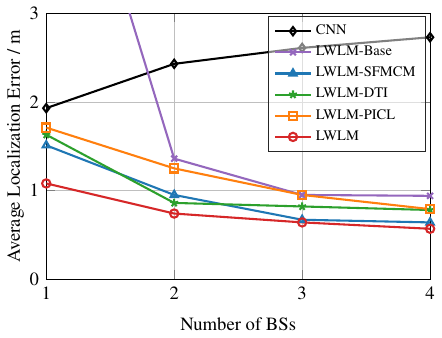}
    \vspace{-3mm}
}

\vspace{-2mm}

\subfloat[\footnotesize Average localization error results under different fine-tuning dataset sizes for 4-BS localization.\label{fig:MB_dataset}]{
    \includegraphics[width=0.37\textwidth]{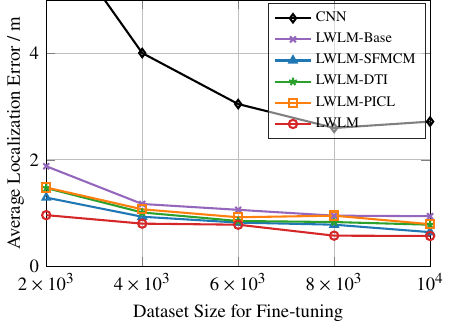}
}

\caption{\footnotesize Performance evaluation of multi-BS localization tasks under different settings.}
\label{fig:MB_combined}
\end{figure}

Fig.~\ref{fig:MB_combined}\subref{fig:MB_result_CDF} illustrates the CDF of localization errors with 4 BSs and a fine-tuning dataset of 10,000 samples. As shown in Fig.~\ref{fig:MB_combined}\subref{fig:MB_result_CDF}, even without any pretraining, the transformer-based LWLM-Base model achieves a 50\% localization error of approximately 0.74 m, significantly outperforming the CNN-based baseline. This performance gain stems from the multi-BS setting, where the model benefits from learning position-invariant semantics across different BSs, thus facilitating better generalization. When a pretraining strategy is adopted, the localization performance is further improved. Among the single-pretraining approaches (i.e., LWLM-SFMCM, LWLM-DTI, and LWLM-PICL), each contributes to different degrees of accuracy enhancement. 
Our proposed LWLM model, which integrates all three SSL strategies in a hybrid manner, achieves the best performance with a 50\% localization error of only 0.51 m. Compared to models using a single pretraining strategy, this result corresponds to a relative improvement ranging from 1.9\% to 25.9\%. Furthermore, LWLM achieves a 32.0\% improvement in localization accuracy over LWLM-Base, which does not use any pretraining. These results clearly demonstrate the advantage of hybrid pretraining in enhancing generalization for multi-BS localization.

Fig.~\ref{fig:MB_combined}\subref{fig:MB_basestations} illustrates the impact of the number of BSs on localization performance. As shown in Fig.~\ref{fig:MB_combined}\subref{fig:MB_basestations}, the localization error of all transformer-based methods consistently decreases as the number of BSs increases. This is expected, as more BSs provide richer spatial information, which helps the model better disambiguate UE positions and improves estimation certainty. In contrast, the CNN method exhibits a performance degradation when the number of BSs increases. This is because the increase in the number of BSs leads to an increase in the input dimension, which exacerbates the overfitting of the CNN architecture. However, CNN is unable to perform effective semantic aggregation, which increases the localization error during testing. Among all methods, the proposed hybrid pretraining model LWLM consistently achieves the best localization accuracy. When using 4 BSs, LWLM achieves an average localization error of 0.57 meters, which is 11.9\% to 27.8\% higher than all single pretraining algorithms. This highlights the robustness and scalability of the hybrid self-supervised framework in multi-BS scenarios.

Fig.~\ref{fig:MB_combined}\subref{fig:MB_dataset} shows the average localization error under different fine-tuning dataset sizes. Increasing the amount of labeled data improves the performance of all models. Our proposed LWLM consistently outperforms all baselines across all data scales. Remarkably, even with only 2,000 fine-tuning samples, LWLM achieves an average localization error of 0.96~m, which represents a relative improvement of 25.6\%–48.9\% over other baselines based on the transformer. This demonstrates the strong data efficiency and few-shot generalization capability of LWLM in multi-BS localization scenarios.

\section{Conclusion}
\label{sec:Conclusion}

In this paper, we proposed the LWLM, a transformer-based self-supervised foundation model tailored for wireless localization tasks. To overcome the limitations of task-specific models and improve generalization across diverse wireless configurations, we introduced a hybrid SSL framework that integrates SF-MCM, DTI, and PICL. These objectives enable LWLM to learn rich, diverse, and transferable channel semantics from large-scale unlabeled data. We further designed lightweight task-specific decoders for a range of downstream localization tasks, including ToA and AoA estimation, as well as single-BS and multi-BS localization. Extensive experiments on simulated datasets demonstrate that LWLM significantly outperforms existing supervised and unsupervised baselines across all tasks. In particular, LWLM achieves superior accuracy with minimal labeled data and exhibits strong robustness to unseen BS configurations. Overall, LWLM represents a scalable and flexible foundation model for wireless localization, capable of supporting future 5G and 6G networks in complex and dynamic environments.

\appendices
\section{Proof of Theorem 1}
\label{proof:theorem1}
For each generative-based SSL task, the IB objective can be expanded as $I(O; H) - \beta_{k_g} I\left(O; T_{k_g}(H)\right) = H(H)\!- \! H(H|O) \!- \! \beta_{k_g} \left(H(T_{k_g}(H)) \!- \! H(T_{k_g}(H)|O)\right) = H(H)- H\left(T_{k_g}(H)|O \right) + H\left(\overline{T}_{k_g}(H)|T_{k_g}(H),O\right)  - \beta_{k_g} \left(H(T_{k_g}(H)) -H(T_{k_g}(H)|O)\right)  = H(H)- \beta_{k_g} H\left(T_{k_g}(H)\right) + (\beta_{k_g}-1) H\left(T_{k_g}(H)|O\right) - H\left(\overline{T}_{k_g}(H)|T_{k_g}(H),O\right)$. Here, $H(H)- \beta_{k_g} H\left(T_{k_g}(H)\right) = C$ is a constant independent of the parameters $\bm{\Phi}$, $\bm{\Theta}_{k_g}$ and $\overline{\bm{\Theta}}_{k_g}$.

We assume that $q_{\bm{\Phi}}(O|H) \approx q(O|H) $, $p_{\bm{\Theta}_{k_g}}(T_{k_g}(H)|O) \approx p(T_{k_g}(H)|O) $, and $p_{\overline{\bm{\Theta}}_{k_g}}(\overline{T}_{k_g}(H)|T_{k_g}(H),O) \approx p(\overline{T}_{k_g}(H)|T_{k_g}(H),O)$. \footnote{In practice, we assume that the model capacity is large enough for these parameterized distributions to closely match the true ones.} Generative model $p_{\bm{\Theta}_{k_g}}(T_{k_g}(H)|O)$ and $p_{\overline{\bm{\Theta}}_{k_g}}(\overline{T}_{k_g}(H)|O)$ can also be considered as decoders for $T_{k_g}(H)$ and $\overline{T}_{k_g}(H)$. Then, we have 
\begin{align}
\! \! \! &H\left(T_{k_g}(H)|O\right) = \mathbb{E}_{p(O,H)}\left[\ln p(T_{k_g}(H)\,|\,O)\right] \nonumber \\
\! \! \!\approx& \mathbb{E}_{p(H)q_{\bm{\Phi}}(O|H)} \! \! \left[\ln  p_{\bm{\Theta}_{k_g}}\!\! \left(T_{k_g}(H)|O\right)\right] \! = \!  -\mathcal{L}_{k_g}(\bm{\Phi},\! \bm{\Theta}_{k_g}),   
\end{align}
and 
\begin{align}
& H\! \! \left(\overline{T}_{k_g}\! (H)|T_{k_g}(H),\! O\right) \!  = \!  \mathbb{E}_{p(O,H)}\! \! \left[\ln p\! \left(\overline{T}_{k_g}(H)|T_{k_g}(H),\! O\right)\right] \nonumber\\ 
 \approx  &
\mathbb{E}_{p(H)q_{\bm{\Phi}}(O|H)}\left[\ln p_{\overline{\bm{\Theta}}_{k_g}}\left(\overline{T}_{k_g}(H)|T_{k_g}(H),O\right)\right]  \nonumber\\ 
= & -\overline{\mathcal{L}}_{k_g}(\bm{\Phi},\overline{\bm{\Theta}}_{k_g}).
\end{align}
Based on above, We have $I(O;\! H) \!- \! \beta_{k_g} \! I \! \left(O; T_{k_g}(H)\right) \! \approx \! (\beta_{k_g} \!- \! 1) \mathcal{L}_{k_g}(\bm{\Phi},\bm{\Theta}_{k_g}) -  \overline{\mathcal{L}}_{k_g}(\bm{\Phi},\overline{\bm{\Theta}}_{k_g}) + C$, which completes the proof.

\section{Proof of Theorem 2}
\label{proof:theorem2}

Let $O$ and $O^+$ denote the random variables corresponding to the encoder representations of two channel observations $(\bm{H}_s, \bm{H}_{s^+})$ that share the same label $\bm{y}_s$. Assuming the encoder operates independently across samples, and both $\bm{H}_s$ and $\bm{H}_{s^+}$ are drawn conditionally on the same label, we have the Markov chain of random variables:
$O \rightarrow Y \rightarrow O^+$.
By the data processing inequality, we obtain
 $   I(O; Y) \ge I(O; O^+)$.
%
Following the result from~\cite{oord2018representation}, the InfoNCE loss $L_{\text{InfoNCE}}$ provides a variational lower bound on $I(O; O^+)$:
    $I(O; O^+) \ge \log(N_{\text{bat}}) - L_{\text{InfoNCE}}$. 
%
Combining the two inequalities, we get
    $I(O; Y) \ge I( O; O^+ ) \ge \log(N_{\text{bat}}) - L_{\text{InfoNCE}}$,
which completes the proof.

\bibliographystyle{IEEEtran}
\bibliography{IEEEabrv, ref}

\newpage

\end{document}